%% file: sample631.tex
\documentclass[twocolumn, twocolappendix]{aastex702}

\usepackage{amsmath,amsthm} 
\usepackage{amsfonts} 
\usepackage{graphicx} 
\usepackage{multirow}

\usepackage{xcolor}
\usepackage{amsmath}
\usepackage{CJK}
\usepackage{changepage}

\definecolor{Goldenrod}{RGB}{255, 215, 0}

\newcommand{\gold}{\textcolor{Goldenrod}{G}}
\newcommand{\silver}{\textcolor{gray}{S}}
\newcommand{\bronze}{\textcolor{brown}{B}}

\newcommand{\hbeta}{H{$\beta$}}
\newcommand{\halpha}{H{$\alpha$}}

\newcommand{\CIV}{C{\sevenrm\,IV}}

\def \OIII {[O\,{\sc iii}]}

\newcommand{\MgII}{Mg{\sevenrm\,II}}

\newcommand{\bracket}[1]{\left\langle#1\right\rangle}
   \font\sevenrm=cmr7 scaled 1000

\newcommand{\mbh}{$M_{\rm BH}$}
\defcitealias{ShenEtAl2024}{S24}
\defcitealias{MacDougallEtAl2026}{M26}

\graphicspath{{./}{figures/}}
\begin{document}

\title{The SDSS-V Black Hole Mapper Reverberation Mapping Project: Short Lags in the COSMOS Field}

\author[0000-0002-8501-3518]{Zachary Stone}
\affiliation{Department of Astronomy, University of Illinois at Urbana-Champaign, Urbana, IL 61801, USA}
\email[show]{stone28@illinois.edu}

\author[0000-0003-1659-7035]{Yue Shen}
\affiliation{Department of Astronomy, University of Illinois at Urbana-Champaign, Urbana, IL 61801, USA}
\affiliation{National Center for Supercomputing Applications, University of Illinois at Urbana-Champaign, Urbana, IL 61801, USA}
\email{shenyue@illinois.edu}

\author[0000-0002-6404-9562]{Scott F. Anderson}
\affiliation{Department of Astronomy, University of Washington, Box 351580, Seattle, WA 98195, USA}
\email{sfander@uw.edu}

\author[0000-0002-9508-3667]{Roberto Assef}
\affiliation{N\'ucleo de Astronom\'ia de la Facultad de Ingenier\'ia y Ciencias, Universidad Diego Portales, Av. Ej\'ercito Libertador 441, Santiago, Chile}
\email{roberto.assef@mail.udp.cl}

\author[0000-0002-8686-8737]{Franz E. Bauer}
\affiliation{Instituto de Astrof\'isica, Pontificia Universidad Cat\'olica de Chile, Av. Vicuna Mackenna 4860, 782-0436 Macul, Santiago, Chile}
\email{fbauer@astro.puc.cl}

\author[0000-0002-0167-2453]{W. N. Brandt}
\affiliation{Department of Astronomy \& Astrophysics, 525 Davey Lab, The Pennsylvania State University, University Park, PA 16802, USA}
\affiliation{Institute for Gravitation and the Cosmos, Pennsylvania State University, University Park, PA 16802, USA}
\affiliation{Department of Physics, 104 Davey Lab, The Pennsylvania State University, University Park, PA 16802, USA}
\email{wnb3@psu.edu}

\author[0000-0002-8725-1069]{Joel R. Brownstein}
\affiliation{Department of Physics and Astronomy, University of Utah, 270 S. 1400 E. \#E2108, Salt Lake City, UT 84112, USA}
\email{joelbrownstein@astro.utah.edu}

% \author[0000-0001-9831-7775]{Vivian Carvajal}
% \affiliation{Steward Observatory, University of Arizona, Tucson, AZ 85721, USA}
% \email{vcarvajal@arizona.edu}

\author[0000-0001-9776-9227]{Megan C. Davis}
\affiliation{Department of Physics, University of Connecticut, 2152 Hillside Road, Unit 3046, Storrs, CT 06269, USA}
\email{megan.c.davis@uconn.edu}

\author[0000-0003-3903-0373]{Jason Dexter}
\affiliation{JILA, University of Colorado and National Institute of Standards and Technology, 440 UCB, Boulder, CO 80309-0440, USA}
\affiliation{Department of Astrophysical and Planetary Sciences, University of Colorado, 391 UCB, Boulder, CO 80309-0391, USA}
\email{Jason.Dexter@colorado.edu}

\author[0000-0002-4459-9233]{Tom Dwelly}
\affiliation{Max-Planck-Institut f\"ur Extraterrestrische Physik
Giessenbachstra{\ss}e, 85748 Garching, Germany}
\email{dwelly@mpe.mpg.de}

\author[0000-0003-3310-0131]{Xiaohui Fan}
\affiliation{Steward Observatory, University of Arizona, Tucson, AZ 85721, USA}
\email{fan@as.arizona.edu}

\author[0000-0001-9920-6057]{Catherine J. Grier}
\affiliation{Department of Astronomy, University of Wisconsin-Madison, 475 N. Charter St, Madison, WI 53706, USA} 
\email{kate.grier@astro.wisc.edu}

\author[0000-0001-9440-8872]{Norman A. Grogin}
\affiliation{Space Telescope Science Institute, 3700 San Martin Drive, Baltimore, MD 21218, USA}
\email{nagrogin@stsci.edu}

\author[0000-0002-1763-5825]{P. B. Hall}
\affiliation{Department of Physics and Astronomy, York University, Toronto, ON M3J 1P3, Canada}
\email{phall@yorku.ca}

\author[0000-0002-8606-6961]{Lorena Hern\'andez-Garc\'ia}
\affiliation{Universidad Diego Portales, Instituto de Estudios Astrof\'isicos, Facultad de Ingenier\'ia y Ciencias, Av. Ej\'ercito Libertador 441, Santiago, Chile}
\affiliation{Centro Interdisciplinario de Data Science, Facultad de Ingenier\'ia y Ciencias, Universidad Diego Portales, Av. Ej\'ercito Libertador 441, Santiago, Chile}
\email{lorena.hernandez@mail.udp.cl}

\author[0000-0003-1728-0304]{Keith Horne}
\affiliation{University of St Andrews, SUPA School of Physics \& Astronomy, North Haugh, St Andrews KY16 9SS, UK}
\email{kdh1@st-andrews.ac.uk}

\author[0000-0002-0957-7151]{Yasaman Homayouni}
\affiliation{Department of Astronomy \& Astrophysics, 525 Davey Lab, The Pennsylvania State University, University Park, PA 16802, USA}
\affiliation{Institute for Gravitation and the Cosmos, Pennsylvania State University, University Park, PA 16802, USA}
\email{ybh5251@psu.edu}

\author[0000-0002-6610-2048]{Anton M. Koekemoer}
\affiliation{Space Telescope Science Institute, 3700 San Martin Drive, Baltimore, MD 21218, USA}
\email{koekemoer@stsci.edu}

\author[0000-0002-7955-7359]{Dhanesh Krishnarao}
\affiliation{Department of Physics, Colorado College, 14 East Cache la Poudre St., Colorado Springs, CO, 80903, USA}
\email{dkrishnarao@coloradocollege.edu}

\author[0000-0002-0311-2812]{Jennifer Li}
\affiliation{National Center for Supercomputing Applications, University of Illinois at Urbana-Champaign, Urbana, IL 61801, USA}
\email{jli184@illinois.edu}

\author[]{Diego Mardones-Soto}
\affiliation{Departamento de Astronom\'ia, Universidad de Concepci\'on, Casilla 160-C, Concepci\'on, Chile}
\email{diemardones2021@udec.cl}

\author[0000-0002-7843-7689]{Mary Loli Mart\'inez-Aldama}
\affiliation{Universidad de Chile, Av. Libertador Bernardo O'Higgins 1058, Santiago de Chile}
\email{mmartinez@das.uchile.cl}

\author[0000-0002-4134-864X]{Knox Long}
\affiliation{Space Telescope Science Institute, 3700 San Martin Drive, Baltimore, MD 21218, USA}
\email{long@stsci.edu}

\author[0000-0002-6770-2627]{Sean Morrison}
\affiliation{Department of Astronomy, University of Illinois at Urbana-Champaign, Urbana, IL 61801, USA}
\email{smorris0@illinois.edu}

\author[]{Leonardo Olivares Latorre}
\affiliation{Departamento de Astronom\'ia, Universidad de Concepci\'on, Casilla 160-C, Concepci\'on, Chile}
\email{lolivares2022@udec.cl}

\author[orcid=0000-0002-0033-5041]{Theodore Pena}
\affiliation{Department of Astronomy, University of Wisconsin-Madison, 475 N. Charter St, Madison, WI 53706, USA} 
\email{theodore.pena@wisc.edu}

\author[0000-0001-6481-5397]{B.~M.~Peterson}
\affiliation{Retired}
\email{}

% \author[0000-0001-5231-2645]{Claudio Ricci}
% \affiliation{Instituto de Estudios Astrofísicos, Facultad de Ingenier\'ia y Ciencias, Universidad Diego Portales, Avenida Ejército Libertador 441, Santiago, Chile}
% \affiliation{Kavli Institute for Astronomy and Astrophysics, Peking University, Beijing 100871, People’s Republic of China}
% \email{claudio.ricci@mail.udp.cl}

\author[0000-0001-7240-7449]{Donald P. Schneider}
\affiliation{Department of Astronomy \& Astrophysics, 525 Davey Lab, The Pennsylvania State University, University Park, PA 16802, USA}
\affiliation{Institute for Gravitation and the Cosmos, Pennsylvania State University, University Park, PA 16802, USA}
\email{dps7@psu.edu}

\author[0000-0002-3683-7297]{Benny Trakhtenbrot}
\affiliation{School of Physics and Astronomy, Tel Aviv University, Tel Aviv 69978, Israel}
\email{bennyt@tauex.tau.ac.il}

\author[0000-0001-8433-550X]{Matthew J. Temple}
\affiliation{N\'ucleo de Astronom\'ia de la Facultad de Ingenier\'ia y Ciencias, Universidad Diego Portales, Av. Ej\'ercito Libertador 441, Santiago, Chile}
\email{matthew.temple@mail.udp.cl}

\author[0000-0002-1410-0470]{Jonathan R. Trump}
\affiliation{Department of Physics, University of Connecticut, 2152 Hillside Road, Unit 3046, Storrs, CT 06269, USA}
\email{jonathan.trump@uconn.edu}

\author[0000-0002-6893-3742]{Qian Yang}
\affiliation{Center for Astrophysics $|$ Harvard \& Smithsonian, 60 Garden St., Cambridge, MA 02138, USA}
\email{qian.yang@cfa.harvard.edu}

\begin{abstract}
We present the first reverberation mapping (RM) measurements in the COSMOS field included in the SDSS-V Black Hole Mapper Reverberation Mapping Project. Using $\sim$6 months of plate-based spectroscopy (from late 2020 to mid-2021) and accompanying ZTF+DECam \emph{gri}-band photometric light curves, we analyze the variability of over 250 active galactic nuclei (AGNs) out to $z \sim 3$. We successfully recover reliable $\lesssim$ 200d time lags for 22 AGNs ($z \lesssim 2.5$) in \halpha, \hbeta, \MgII, or \CIV, using a combination of multiple lag detection methods and significance tests. Our measurements are consistent with the radius-luminosity relations from recent large-scale RM programs, providing measured masses of $10^{7-9} M_{\odot}$ for these AGNs. These results and upcoming long-baseline SDSS-V spectroscopy lay the groundwork for large-sample AGN studies in conjunction with other facilities covering the COSMOS field, including the Vera C. Rubin Observatory and Roman Space Telescope.
\end{abstract}

% We incorporate LITMUS into our procedure, granting more estimates than with the traditional SDSS-RM pipeline alone. 


%%%%%%%%%%%%%%%%%%%%%%%%%%%%%%%%%%%%%%%%%%%%%%%%%%%%%%%%%%%%%%%%%%%%%%%%%%%%%%%%

\section{Introduction} \label{sec:intro}

%the lamppost model \citep{Fredjung1985, SergeevEtAl2005} where

Active galactic nuclei (AGNs) represent a prime population for constraining supermassive black hole (SMBH) growth across cosmic time. AGN SMBH mass estimation capitalizes on the signature multi-band, multi-timescale variability within AGNs, well-described by Gaussian random processes \citep[e.g.,][]{KellyEtAl2009, KellyEtAl2014, SimmEtAl2016}. In particular, the reverberation mapping \citep[RM;][]{BlandfordMckee1982} method assumes the signal from the central engine close to the SMBH is reverberated throughout the system, with the reprocessing of radiation dominated by light travel time \citep{Peterson2001}. While the origin of said variability remains under debate, possibly originating from instabilities within the accretion disk \cite[e.g.,][]{BalbusHawley1991, JiangBlaes2020, NeustadtKochanek2022, YaoEtAl2023}, estimating a time lag between two regions within the AGN can nonetheless provide an approximate distance. Short time lags between two photometric bands can be used to estimate the size of the accretion disk \citep[e.g.,][]{CollierEtAl1999}, while a lag between the photometric continuum and spectroscopic broad line light curves grant a size for the broad-line region \citep[BLR;][]{Peterson2001}. Assuming a virial BLR, a property which may be violated in certain unique systems \citep[e.g.,][]{DehghanianEtAl2019, StoneEtAl2025}, combining the BLR radius and BLR velocity (estimated from the broad lines), produces a SMBH mass estimate \mbh.

Prior RM campaigns have revealed a tight correlation between the BLR radius and the luminosity of the AGN, dubbed the $R-L$ relation \citep[e.g.,][]{BentzEtAl2013}. Utilizing this correlation enables a \mbh\ estimate from just a single spectrum, providing an estimate of the BLR velocity and $L$ \citep[e.g.,][]{WandelEtAl1999, McLureJarvis2002, Vestergaard2002, VestergaardPeterson2006, Shen2013, DallaBontaEtAl2020}. Therefore, RM campaigns present an avenue to calibrate single-epoch estimator recipes used on millions of AGNs \citep[e.g.,][]{WuShen2022}. Velocity-resolved RM uses high-resolution, high-cadence data to probe the geometry and kinematics of gas within the BLR, able to refine RM estimates by determining how well the virial assumption holds in individual systems \citep[e.g.,][]{GrierEtAl2017_model, VillafanaEtAl2022}. Such analyses estimate virial factors $f$ that modify the virial product:
\begin{equation}
    M_{\rm BH} = f \frac{c\tau \Delta V^2}{G}
\end{equation}
where $\Delta V$ is the estimate of BLR gas velocity inferred by the line width and $\tau$ is the time lag. Dynamical models suggest a population-wide $\bracket{\log_{10}f} = 0.6$, using the emission line velocity dispersion, with $\sim0.3$~dex scatter \citep{ShenEtAl2024}.

To properly sample the months-to-years-long time lags, RM programs require high-cadence, high-resolution, high-quality spectroscopy and accompanying simultaneous photometry. Early RM programs focused on $\sim$ dozens of local ($z < 1$) sources at most \citep[e.g.,][]{BentzEtAl2010}. Recently, industrial-scale survey-based RM programs have been able to measure hundreds of SMBH masses out to $z \lesssim 3$, drastically expanding the sample of \mbh\ estimates \citep[e.g.,][]{ShenEtAl2015, ShenEtAl2024,MacDougallEtAl2025}. Probing such a wide variety of SMBHs across both redshift and AGN parameter-space (e.g., luminosity, accretion rate) is essential to uncovering the mechanisms behind SMBH growth and providing a ground truth for cosmological models. 

The COSMOS field plays a key role in the current time-domain astronomy ecosystem, designated as a Deep-Drilling Field in the Vera C. Rubin Observatory's Legacy Survey of Space and Time \citep[LSST;][]{IvezicEtAl2019} with a few Ultra-Deep seasons. As part of the Nancy Grace Roman Space Telescope's High-Latitude Wide-Area Survey (HLWAS), COSMOS will receive repeated deep near-infrared imaging and spectroscopy. COSMOS-Web \citep{cosmos_webb} and COSMOS-3D \citep{cosmos3d} have been observing COSMOS in the near- and mid-infrared with multiple epochs of imaging and spectroscopy using JWST, discovering dozens of AGNs at $1 \lesssim z \lesssim 3$ \citep{JiangEtAl2026} and hundreds of candidate little red dots \citep{MattheeEtAl2024} at $5 \lesssim z \lesssim 9$ \citep{AkinsEtAl2025}. As one of the most studied fields, COSMOS has a wealth of photometric and spectroscopic data across different regimes, including X-ray \citep[e.g.,][]{cosmos_xray1, cosmos_xray2}, optical \citep[e.g.,][]{cosmos_hsc}, infrared \citep[e.g.,][]{cosmos_ultravista}, sub-mm \citep[e.g.,][]{cosmos_alma}, and radio \citep[e.g.,][]{cosmos_radio}. As such, vast numbers of sources within COSMOS have secure spectroscopic redshifts and classifications \citep[e.g.,][]{cosmos_specz}.

\begin{figure*}
    \centering
    \includegraphics[width=\textwidth]{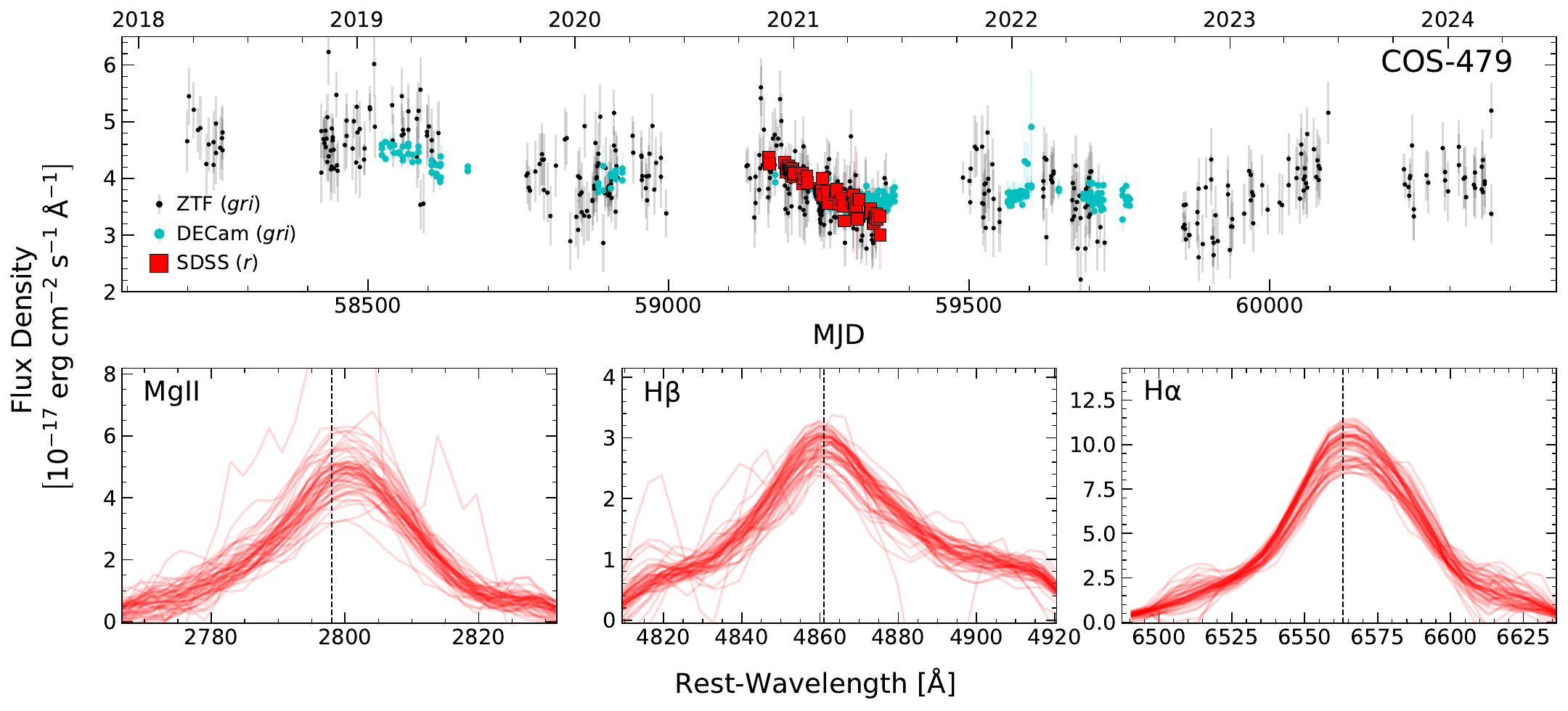}
    \caption{An example RM dataset from the sample. \emph{Top}: Combined multi-band \emph{gri} photometry using PyCALI. Photometry from different facilities is labeled using different symbols. \emph{Bottom}: Multi-epoch broad-line components for three emission lines. Profiles have been smoothed out with a 3-pixel boxcar kernel. A vertical dashed line shows the center of the line.}
    \label{fig:cont_lc}
\end{figure*}

One of the key components of the fifth iteration of the Sloan Digital Sky Survey \citep[SDSS-V; ][]{sdss5} is the Black Hole Mapper (BHM; Trump et al. in prep), placing a focus on continuing and expanding the SDSS-RM program \citep{ShenEtAl2015, ShenEtAl2024}. The SDSS-V BHM-RM project has been amassing repeated spectroscopy of thousands of RM targets across a number of legacy time-domain fields, including COSMOS. Here we present the first RM SMBH mass estimates in COSMOS from the SDSS-V BHM-RM project, utilizing the first six months of spectroscopy to obtain short time lags. In this analysis, we aim to establish the pipeline used to obtain said estimates with the quality and length of current data, expanding our current procedure to incorporate newer methods as well.

In Section~\ref{sec:data}, we describe the spectroscopic and photometric data used, and the sample of broad-line AGNs. We also provide a brief description of our lag-estimation procedure, generally following that of the SDSS-RM project \citep[][hereafter S24]{ShenEtAl2024} and incorporating methods from the more recent OzDES RM program results \citep[][hereafter M26]{MacDougallEtAl2026}. In Section~\ref{sec:results}, we investigate the reliability of our lag estimates, compare methodology between \citetalias{ShenEtAl2024} and \citetalias{MacDougallEtAl2026}, and provide \mbh\, using 28 reliable lags. Finally, we conclude in Section~\ref{sec:conclusion}. Throughout the paper, we adopt a cosmology with $H_0=70$ km s$^{-1}$ Mpc$^{-1}$, $\Omega_{\rm m}=0.3$, and $\Omega_{\Lambda}=0.7$.

\section{Data}\label{sec:data}

\subsection{Spectroscopy}

In late 2020 to mid-2021, continuing from previous iterations of SDSS \citep{SDSS_telescope}, SDSS-V used plates to position fibers for observations. These plates offered the same quality of observations as in SDSS-IV \citep{sdss4}, in terms of signal-to-noise (SNR), astrometry, and spectro-photometric calibration. In late 2021, SDSS-V switched its observing strategy to use a focal plane system (FPS), with robots  controlling the positioning of the fibers instead of the plates \citep{sdss_dr19}. This initial implementation of the FPS faced unique challenges in terms of spectro-photometric calibration, essential for reverberation mapping. There are still unresolved systematics in current FPS spectro-photometry, especially for high-redshift sources without reliable narrow line calibrators (e.g., \OIII). Given the level of uncertainty and still ongoing efforts by the SDSS-V team to mitigate such issues, we opt to use only the plate data in this analysis for the most secure measurements. Future work will use the entire BHM spectroscopic dataset within COSMOS.

We compile 293 broad-line AGNs within the COSMOS field, using the {\tt spAll-master} file (as of early 2024, up to v6.1.1), which aggregates information for all spectroscopic observations. We first restrict to spectroscopy utilizing the plates, occurring over six months from late 2020 to mid-2021 at the Apache Point Observatory (APO) on the Baryon Oscillation Spectroscopic Survey (BOSS) spectrograph \citep{BOSS_spec, sdss_dr19}. Individual spectra within COSMOS are matched to each other to provide a list of spectra for each source, using a matching radius of 0\farcs3 with the provided {\tt RACAT} and {\tt DECCAT}. We visually inspect all spectra to confirm that they are bona-fide broad-line AGNs, identifying either significant continuum/broad-line variability, the presence of typical broad emission lines (e.g., \MgII, \halpha, \hbeta), and/or a strong power-law continuum. We compile an initial sample of 299 vetted AGNs in COSMOS.

We remove spectra with the following bitmask {\tt ZWARNING} flags: {\tt BAD\_TARGET}, {\tt NODATA}, {\tt UNPLUGGED}, {\tt LITTLE\_COVERAGE}. We also restrict to spectra that were flagged as {\tt ON\_TARGET}, {\tt VALID}, {\tt ASSIGNED}, and {\tt DECOLLIDED}. We additionally remove any observations visually inspected to have observational artifacts that were not caught by the BOSS reduction pipeline \citep[][Morrison et al. in prep]{BoltonEtAl2012, DawsonEtAl2013}, totaling 19 spectra across six AGNs. This leaves a final sample of 293 COSMOS AGNs for subsequent analysis. We further inspect the pipeline-reduced redshifts, and correct those that are discrepant by using prominent emission lines if available (e.g., \OIII), or a combination of broad lines if not. Only one object was assigned an incorrect redshift (COS-934), which is not included in our final analysis. Our chosen spectroscopy has an average SNR {\tt SN\_MEDIAN\_ALL}$\sim7$ and mean cadence $\approx$3.3d over $\approx55$ epochs from November 2020 to May 2021. 20/293 AGNs have $<30$ epochs, only five of which were included in our final analysis (COS-608, COS-1342, COS-1380, COS-1552, COS-1554), and none of which are included in our final lag sample.

To obtain reliable emission-line measurements and spectral information, we use PrepSpec \citep[see][]{ShenEtAl2016} to prepare the spectra. Briefly, PrepSpec iteratively fits a flexible spline-based model, including the continuum and broad+narrow emission lines, to all of the spectra for each object, calibrating with narrow emission lines. PrepSpec provides broad emission line light curves for all significant broad lines, as well as broad emission-line widths and spectral continuum measurements. We utilize PrepSpec to measure \halpha, \hbeta, \MgII, and \CIV\, -- emission lines common in optical spectroscopy and that have shown reverberations relative to the continuum \citepalias[e.g.,][]{ShenEtAl2024}.

 % This leaves a sample of 261 AGNs in the COSMOS field.

% \begin{itemize}
%     \item During initial data acquisition with the fiber positioning system (FPS) in late-2021, spectrophotometry was poorer than expected. While the spectrophotometry has improved in the meantime, there continue to be some issues with spectroscopic calibration. While these issues are being iterated upon, we restrict this analysis to plate data for the most reliable and significant RM lag estimates.
%     \item Query for all plate-era spectra in the SpAll file in the COSMOS field
%     \item Aggregate spectra from individual sources by matching with {\tt RACAT} and {\tt DECCAT} with a matching radius of \blue{xxx}
%     \item Remove spectra with bad flags or visible artifacts
%     \item Visually inspect all sources to confirm if they are broad-line AGNs
%     \item Visually inspect pipeline-produced redshifts, to see if they are correct
%     \item Left with a sample of 261 AGNs (485 potential lags)
% \end{itemize}

\subsection{Photometry}

COSMOS is a popular legacy field for repeated imaging, monitored in \emph{gri} by both the Zwicky Transient Facility \citep[ZTF;][]{ztf} since 2018, and DECam since early 2019 \citep{helm}. For a symmetric photometric baseline about the spectroscopy, we query for all ZTF \emph{gri} photometry (matched within 1") up to early 2024. Specifically, we use a radial search to query the epochal photometric catalogs with ztfquery \citep{ztfquery}, filtering with {\tt catflags = 0} to remove poor quality data. We include ZTF data across fields, as visual inspection verified no noticeable systematics between fields within scatter. We also match to all \emph{gri} DECam light curves from the HELM program up to mid-2022 \citep{helm}, utilizing data with {\tt flags < 4}. While all AGNs were provided photometry by ZTF, 12/293 did not have a match in HELM.

For spectro-photometric calibration, we use SDSS synthetic \emph{r}-band photometry provided in the {\tt spAll} file. All light curves are combined and scaled to SDSS-\emph{r} using PyCALI \citep{pycali}, assuming a Damped Random Walk (DRW) model to find optimal scaling parameters between bands and observatories. We opt to use the same PyCALI sampling parameters as \citetalias{ShenEtAl2024}. Calibrations do not consider short inter-band lags ($\sim$days), which are subdominant to the weeks-to-months-long lags from broad-line reverberation.

Restricting to objects with both spectroscopy and photometry leaves a sample of 261 AGNs at $z \lesssim 3$, with 485 potential time lags to be probed across the four considered lines. All continuum light curves have $\gtrsim 60$ epochs over six years, with some having nearly 1000. To increase the SNR and reduce computational cost, we combine nightly epochs within the photometric light curves, weighting by inverse variance. An example continuum light curve and multi-epoch spectroscopy are shown in Fig~\ref{fig:cont_lc}.

% \begin{figure}
%     \centering
%     \includegraphics[width=\linewidth]{figures/sig0.pdf}
%     \caption{Maximum likelihood intrinsic variability estimates $\sigma_0$ for the entire sample, relative to the mean of the light curve. This includes the variability of both the continuum and emission line light curves, with each panel representing a different line.}
%     \label{fig:sig0}
% \end{figure}

\subsection{Estimating Time Lags with PyPETaL}\label{subsec:lag_methods}

\begin{figure}
    \centering
    \includegraphics[width=\linewidth]{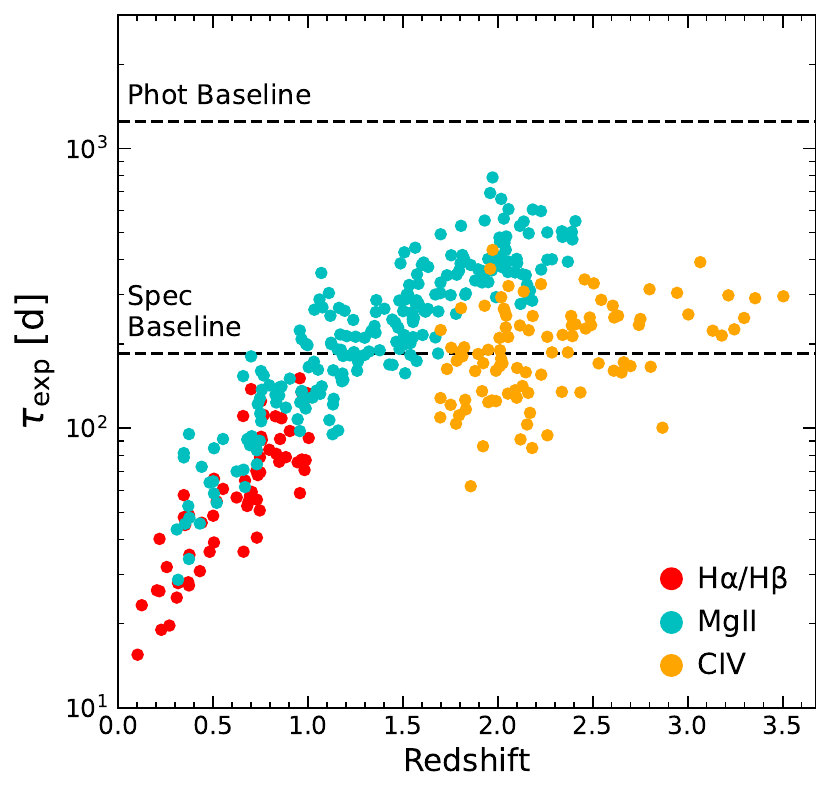}
    \caption{Expected observed lag as a function of redshift for each measured line for the entire sample, using literature $R-L$ relations for each line \citep{BentzEtAl2013, MacDougallEtAl2026}. Each lag is color-coded by its emission line. The maximum photometric and spectroscopic temporal baselines of our sample are displayed as horizontal dashed lines.}
    \label{fig:exp_lag}
\end{figure}
% \blue{Change y-label to be $\tau_{\rm exp}$.}

% \input{figset}
% \begin{figure*}[t]
%     \centering
%     \includegraphics[width=\textwidth]{summary_BHMRM00849_hb.pdf}
%     \caption{An example \hbeta\, lag recovery for COS-849 ($z=0.37$). The continuum (\emph{top}) and \hbeta\, (\emph{bottom}) light curves are shown on the left, alongside the best-fit PyROA model. The best-fit model is shown as a line, with the uncertainties shaded around it. Posterior lag distributions from multiple methods are shown on the right. The top two panels show the ICCF (black) and posterior weights (blue, gray, red). A vertical black dashed line represents the measured median time lag, with the 16th-84th percentile shaded in gray. The expected lag is shown as a red dashed vertical line, and the bounds for the lag search ($\pm3\,\tau_{\rm exp}$) are shown as gray dashed vertical lines. The complete figure set featuring our final set of PyROA lags (20 images) is available in the online journal. \digitalasset}
%     \label{fig:example_lag}
% \end{figure*}

\begin{figure*}[t]
    \centering
    \includegraphics[width=\textwidth]{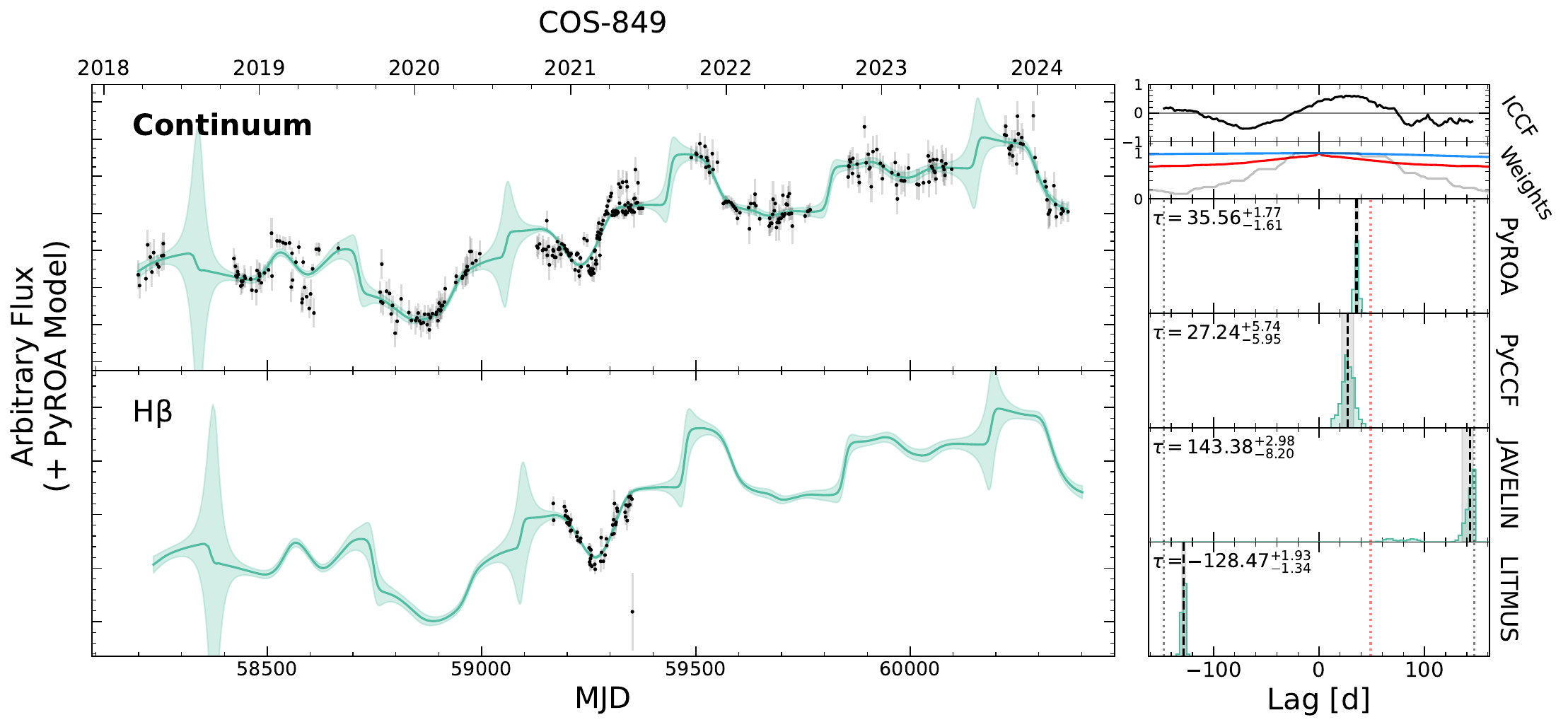}
    \caption{An example \hbeta\, lag recovery for COS-849 ($z=0.37$). The continuum (\emph{top}) and \hbeta\, (\emph{bottom}) light curves are shown on the left, alongside the best-fit PyROA model. The best-fit model is shown as a line, with the uncertainties shaded around it. Posterior lag distributions from multiple methods are shown on the right. The top two panels show the ICCF (black) and posterior weights (blue, gray, red). A vertical black dashed line represents the measured median time lag, with the 16th-84th percentile shaded in gray. The expected lag is shown as a red dashed vertical line, and the bounds for the lag search ($\pm3\,\tau_{\rm exp}$) are shown as gray dashed vertical lines. The complete figure set featuring our final set of PyROA lags (20 images) will be available after publication.}
    \label{fig:example_lag}
\end{figure*}

Generally, we follow a similar methodology to \citetalias{ShenEtAl2024}, shown to work for the SDSS-RM plate-based data. Our time-lag estimation uses PyPETaL \citep{pypetal}, a pipeline combining a number of popular lag-estimation codes: PyCCF \citep{pyccf}, JAVELIN \citep{javelin1, javelin2}, PyROA \citep{pyroa}. To start, we determine a range of lags to search using literature $R-L$ relations. We adopt the \citet{BentzEtAl2013} result for \halpha\ and \hbeta, and the most recent OzDES result for \MgII\, and \CIV\, \citep{MacDougallEtAl2025}. The OzDES $R-L$ relations combine lag measurements from OzDES, SDSS-RM, and prior literature work, giving the most stringent constraints. This is especially true for \CIV, where initial results \citep{KaspiEtAl2021} differ from more recent ones \citepalias{ShenEtAl2024}. We choose a search range of $\pm 3\, \tau_{\rm exp}$, where $\tau_{\rm exp}$ is the expected observed-frame time lag from the chosen $R-L$ relation, utilizing the source's redshift and continuum luminosity $\lambda L_{\lambda}$ (Fig.~\ref{fig:exp_lag}). A range $\pm 3\, \tau_{\rm exp}$ encloses the innate scatter in these relations of $\sim$0.3~dex\footnote{We acknowledge that this choice may preferentially select lags within the scatter of the considered $R-L$ relations. Longer datasets are needed to overcome such bias. }. We reduce the range relative to \citetalias{ShenEtAl2024}, due to the small spectroscopic baseline of our data relative to the photometric baseline. The larger we set the lag search range, the more likely our results will be affected by aliasing from seasonal gaps. For the most secure results, we attempt to avoid this aliasing as much as possible. As such, we consider estimates for sources where $\tau_{\rm exp}$ is less than the spectroscopic baseline the most reliable (see Fig~\ref{fig:exp_lag}). Future measurements with longer spectroscopic baselines will utilize larger lag search ranges. 

We emphasize that our nominal PyPETaL lags assume a symmetric lag range, as opposed to a positive-only search range \citep[e.g.,][]{MacDougallEtAl2025, YuEtAl2023}, to reduce bias towards positive lags. While positive lags are physically-informed, imposing a positive-only prior biases results towards aliased lags within seasonal gaps. 
% Solely for comparison with \citetalias{MacDougallEtAl2026} (Section~\ref{subsec:litmus_fpr}), we also recover lags with a positive-only search range [0, $3\,\tau_{\rm exp}$]. 
% \red{Here we should emphasize the importance of symmetric lag search ranges, with both negative and positive lags. I don't think OzDES did that. }

% and \citet{MacDougallEtAl2025} \red{I thought MacDougal et al. didn't search for negative lags?}

For a complete description of PyPETaL, see \citetalias{ShenEtAl2024}. Briefly, we first fit the continuum light curve to a DRW, and then remove all $5\sigma$ outliers with respect to the model. Afterward, we acquire lag estimates from PyCCF, JAVELIN, and PyROA (Fig.~\ref{fig:example_lag}). Parameters for each of these methods are chosen to properly sample the lag posterior distribution: 5000 samples for PyCCF, $n_{\rm burn} = n_{\rm chain} = n_{\rm walker} = 300$ for JAVELIN, and 15000 samples for PyROA, discarding the first 10000 samples as burn-in. Afterward, we weight the posterior in the same manner as \citet{GrierEtAl2017}, choosing a primary peak after accounting for seasonal gaps. All reported lags use the median and 16th-84th percentiles of this chosen peak for the lag and its lower/upper uncertainties, respectively.

\subsection{LITMUS}

LITMUS \citep{litmus} has recently been developed specifically for estimating RM time lags, using a Bayesian framework to further mitigate the effect of seasonal gaps on lag recoveries (see Section~\ref{subsec:litmus_fpr}). LITMUS adopts a flexible framework, including a number of different Bayesian posterior sampling methods \citep[e.g., nested sampling;][]{NestedSampling} and models. LITMUS was recently used on the entire OzDES RM sample to produce its final set of \mbh\ estimates \citepalias{MacDougallEtAl2026}. For an initial comparison of LITMUS to the \citetalias{ShenEtAl2024} lag-estimation procedure, we procure time lags using LITMUS. We restrict to time lags with $\tau_{\rm exp}$ less than the spectroscopic baseline, so as to compare results for the most reliable lags. We follow \citetalias{MacDougallEtAl2026}, choosing the {\tt GP\_simple} model and using nested sampling, with a maximum 50000 samples and sampling until the Bayesian evidence $Z$ reaches an uncertainty $\Delta Z/Z = 10^{-3}$. This model is similar to JAVELIN's, assuming a DRW for the continuum light curve, and a scaled/delayed continuum for the emission-line response. LITMUS also provides a direct analog to JAVELIN with a tophat response function, although \citetalias{MacDougallEtAl2026} generally recommend the aforementioned simple gaussian process (GP) model instead. Future work will determine how this tophat model compares to JAVELIN, especially for SDSS-V data. We use both a symmetric and positive-only lag search range (bounded at $3\,\tau_{\rm exp}$), to compare with \citetalias{MacDougallEtAl2026} who utilize a positive search range for OzDES.
% \red{(did you still allow negative lags in the search?)}

We extract a posterior with 10000 samples for each line. LITMUS has the advantage of providing a Bayesian evidence for the model fit, allowing us to acquire a Bayes factor (BF) using a model assuming a ``coupled" continuum and emission line (i.e., a lag exists and is present), and a model assuming an ``uncoupled" set of light curves (i.e., no lag exists). As defined in \citetalias{MacDougallEtAl2026}, ${\rm BF_{lag}} = Z_{\rm coupled} / Z_{\rm uncoupled}$ provides a measure for how well a set of light curves encodes a reverberated lag. $\rm BF_{lag} \gg 1$ suggests that a lag exists and is well-measured, while $\rm BF_{lag} \ll 1$ suggests either no lag exists or it is not represented well in the data. $\rm BF_{lag}$ is thus an objective way to probe for false-positives. We acquire lag posteriors and $\rm BF_{lag}$ for all 184 lags with low $\tau_{\rm exp}$. These posteriors are again weighted in the same manner as the other posteriors from PyPETaL.

% \begin{itemize}
%     \item Use {\tt PyPETAL}
%     \item Use a lag range of [$-3\tau_{\rm exp}$, $3\tau_{\rm exp}$]
%     \item Discard $5\sigma$ outliers from continuum DRW fitting
%     \item 5,000 MC samples for {\tt PyCCF}
%     \item $n_{\rm burn} = n_{\rm chain} = n_{\rm walker} = 300$ for {\tt JAVELIN}. Fix the best-fit $\sigma_{\rm DRW},\tau_{\rm DRW}$ from the DRW-fitting
%     \item For PyROA $n_{\rm samp}$ = 15,000  and discard the first 10,000 samples as burn-in
%     \item Use nested sampling from {\tt LITMUS} with the {\tt GP\_simple} model, with a maximum of 50,000 samples, and finishing after reaching an uncertainty in the evidence $\Delta Z/Z = 10^{-3}$. Use best-fit solution to get 10,000 posterior samples
%     \item For all methods get the posterior-weighted lag and $r_{\rm max}$. For {\tt LITMUS}, get the Bayes factor between a coupled response and uncoupled response model ${\rm BF_{lag}}$, following \citet{MacDougallEtAl2026}.
%     \item Restrict to lags following criteria in \citet{ShenEtAl2024}
%     \item Use the procedure in \citet{ShenEtAl2024} to get a false positive rate (FPR) for the lags with {\tt PyCCF}.
%     \item Get $M_{\rm BH}$
% \end{itemize}

\section{Results}\label{sec:results}

\subsection{Lag Reliability}\label{subsec:lag_reliability}

Here, we investigate the quality of our lag estimates. See Appendix~\ref{app:reliability} for an in-depth description of the sample-wide lag quality. We utilize the same criteria as \citetalias{ShenEtAl2024} for selecting an initial sample of reliable lags from our dataset. We choose PyROA as the fiducial lag estimate, as it is much less affected by aliasing than JAVELIN and PyCCF \citepalias[][also see Appendix~\ref{app:lag_comp}]{ShenEtAl2024}. Eventually, we incorporate LITMUS for additional reliable estimates (Section~\ref{subsec:litmus_comp}). One of the metrics we use to determine the reliability of our lags is $r_{\rm max}$, the maximum value of the interpolated cross-correlation function (ICCF) within $\pm1\sigma$ of the best-fit lag. We first restrict to lags with $r_{\rm max} > 0.4$, shown to work well for SDSS-RM \citepalias{ShenEtAl2024}, and those that are positive at the $2\sigma$ level.\footnote{While this may remove the shortest lags, it guarantees the final sample of lags are secure. Nonetheless, a longer dataset is needed to produce uncertainties low enough to recover these lags.} We then reject all lags where more than half of the posterior is rejected from the primary peak selection process. This leaves a sample of 42 lags, with 20 having $\tau_{\rm exp}$ less than the spectroscopic baseline. We restrict all further analysis to these 20 secure lags, to reduce the impact of aliasing on our estimates (see Table~\ref{tab:sample}).

To determine the reliability of our lags, we estimate a false-positive rate (FPR) for each lag, using the same procedure outlined in \citetalias{ShenEtAl2024}. Briefly, we generate 100 realizations of the continuum light curve using the best-fit DRW parameters from PyPETaL. For each realization, we use PyCCF to obtain a best-fit lag with respect to the emission line light curve. The FPR is then the fraction of the 100 lags positive at the $2\sigma$ level with $r_{\rm max} > 0.4$. Like in \citetalias{ShenEtAl2024}, almost all low $\tau_{\rm exp}$ lags have an FPR $< 30\%$, indicating our lags are relatively reliable. The only lag with a poor FPR is the \MgII\, lag in COS-552, likely due to low-SNR spectra contaminating the spectroscopic light curve. We note that even if a PyROA lag estimate is secure, PyCCF may still be susceptible to aliasing and produce a negative lag or $r_{\rm max} < 0.4$. In such cases where the PyCCF lag is not representative of the true lag, the FPR will likely be underestimated, considering the PyCCF lag will not be reliable even with the real data. Thus, low FPRs are expected for \MgII\, and \CIV\, lags with our current dataset, although this can be mitigated using LITMUS' FPR estimation procedure (see Section~\ref{subsec:litmus_fpr}).

\begin{figure}
    \centering
    \includegraphics[width=0.5\textwidth]{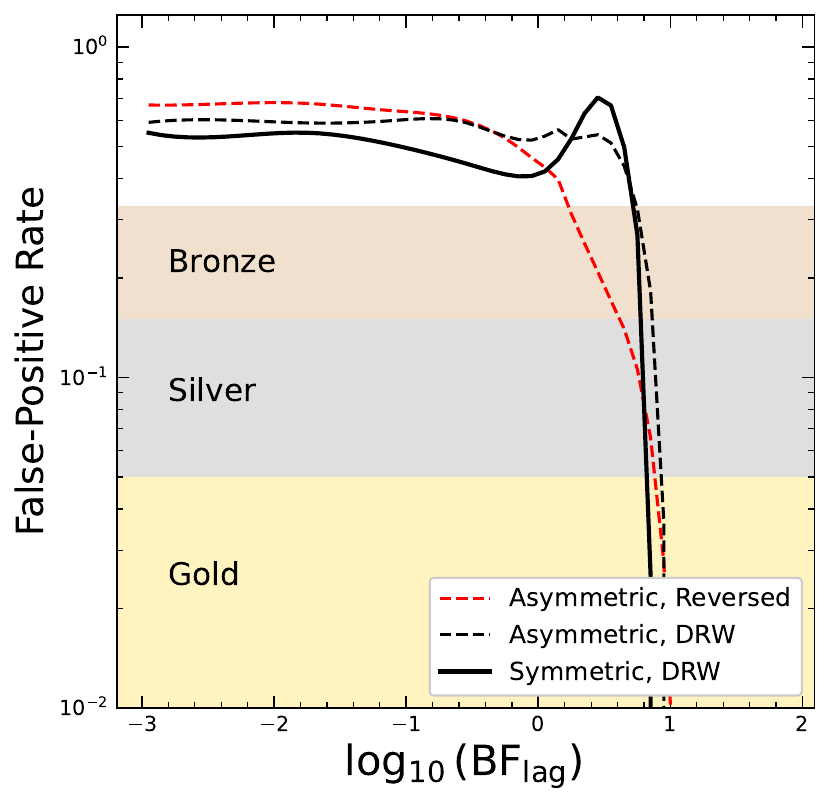}
    \caption{False-positive rate (FPR) curves from LITMUS using three different methods: reversed light curves with an asymmetric search range, random DRW continuum light curve realizations with an asymmetric search range, and the same with a symmetric search range. The shaded regions indicate the ranks given to lags as a function of FPR.}
    \label{fig:fpr_curves}
\end{figure}

Nonetheless, we rank our lag estimates using the five criteria from \citetalias{ShenEtAl2024}. Each criterion satisfied increases a lag's score by one, with gold lags having a score $\geq 4$, silver lags having a score $\in [2,3]$, and bronze lags having the lowest scores. The criteria involve visually inspecting the RMS spectra, the variability of the continuum/line light curves, the posterior lag distribution, and the significance of the lag posterior peak. Out of the 20 lags, 60\% (12) are ranked gold and 40\% (8) are ranked silver. As expected, the shortest Balmer lags are the most reliable, displaying significant continuum variability and clear reverberation. Only a few \MgII\, and \CIV\, lags are recovered due to the short spectroscopic baseline, and most are silver ranked. Results using all 42 lags are diluted with lower-quality estimates, where 43\% (18) are gold, 48\% (20) are silver, and 7\% (3) are bronze.
% We recover $\sim$50\% more lags than the SDSS-RM 1-year results \citep{ShenEtAl2016} due to our improved photometric/spectroscopic sampling and robust lag estimation techniques. 
% \red{Add a brief comment on the comparison of SDSS-RM lags using 1-year data (Shen etal. 2016). Given more light curve points, we are recovering more lags per sample size using 1-year data. }

% \noindent
% \blue{Not including these now, but could:} We include the high $\tau_{\rm exp}$ lags as well, but rank all of them bronze regardless of the criteria they satisfy. 

\begin{figure*}
    \centering
    \includegraphics[width=\textwidth]{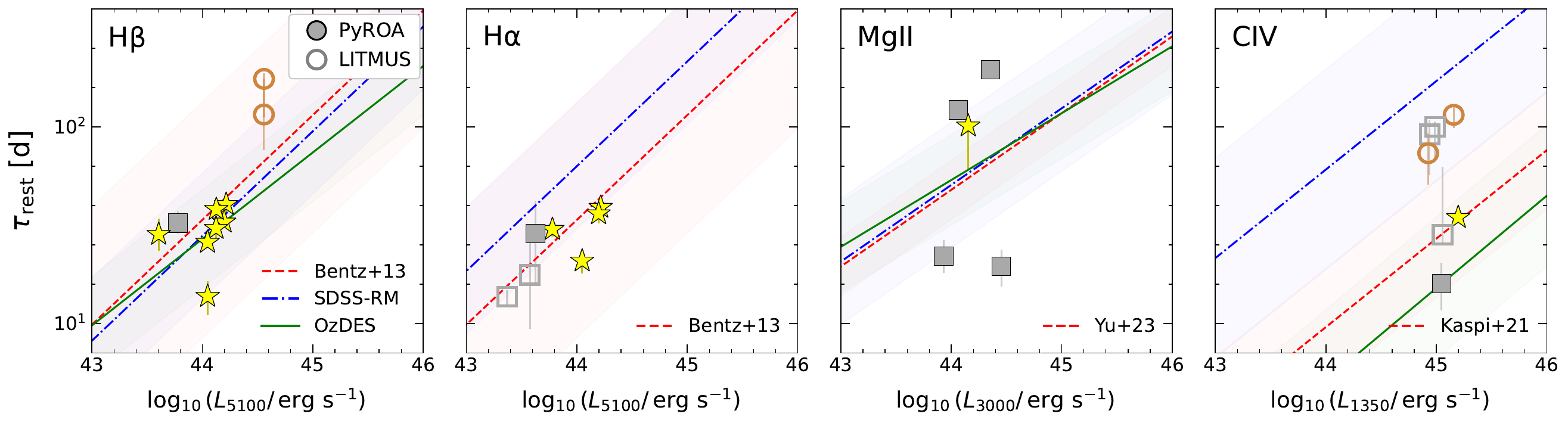}
    \caption{The rest-frame time lags $\tau_{\rm rest}$ measured from our sample, separated by emission line, as a function of continuum luminosity $\lambda L_{\lambda}$. Literature $R-L$ relations, including scatter, are shown for each line \citep{BentzEtAl2013, KaspiEtAl2021, YuEtAl2023, ShenEtAl2024, MacDougallEtAl2025}. The colors and symbols of each estimate correspond to their rank (gold=yellow stars, silver=gray boxes, bronze=brown circles). Lags obtained with PyROA (LITMUS) are shown with filled (open) symbols.}
    \label{fig:rl_relation}
\end{figure*}

\subsection{Bayesian Lag Metrics}\label{subsec:litmus_fpr}

Using LITMUS' Bayesian treatment of lag estimation, we can further define the reliability of our samples' lags. Firstly, we use LITMUS to fit each light curve pair to two additional models: one with a DRW continuum and white-noise response (``GP--noise") and one with a white-noise continuum and a white-noise response (``noise--noise"). The evidence ratio ${\rm BF_{\rm struc,resp}} = Z_{\rm uncoupled}/Z_{\rm GP-noise}$ tells us if the line light curve has significant structure (i.e., variability), and ${\rm BF_{\rm struc,cont}} = Z_{\rm GP-noise}/Z_{\rm noise-noise}$ tells us if the continuum light curve has significant variability. After acquiring posterior distributions for all four models, we implement the same weighting procedure as in Section~\ref{subsec:lag_methods} to find a primary peak in the coupled model. Using the bounds of the peak, we determine the fraction of evidence within the peak. These same bounds in the full coupled model parameter-space are used to obtain the fraction of evidence in the peak for all four considered models, and to adjust the Bayes factors accordingly.

To estimate the FPR for these lags, we follow a similar procedure to Section~4 of \citetalias{MacDougallEtAl2026}. For the lags using a positive-only search range, we fit a reversed version of the two light curves, recovering a lag and all aforementioned Bayes factors. Additionally, we impose that the ``reversed" LITMUS lag meets the three criteria for recovered PyPETaL lags discussed in Section~\ref{subsec:lag_reliability}. For a given $\rm BF_{lag}$ threshold, we record the number of recovered real lags and reversed (i.e., spurious) lags with a higher $\rm BF_{\rm lag}$ (i.e., a cumulative distribution). For each threshold, the ratio of spurious lags to real ones represents the FPR, with bronze lags having FPR $\leq 33\%$, silver lags having FPR $\leq 15\%$, and gold lags having FPR $\leq 5\%$. In other words, if  a given $\rm BF_{lag}$ results in one recovery from the reversed dataset, and 10 recoveries from the real dataset, the FPR is one in 10. To ensure a continuous distribution at all $\rm BF_{lag}$, we fit the cumulative distribution of $\rm \log_{10}(BF_{lag})$ to a SciPy spline (for both the real and spurious lags), and use this model fit to produce an FPR curve (see Fig.~\ref{fig:fpr_curves}). Each lag measurement is assigned an FPR using this FPR curve. For the lags using a symmetric search range, we cannot use the reversed data technique. Instead, we generate a random realization of the continuum light curve using the best-fit model from our initial outlier rejection step, then follow the same previous steps.

We show our FPR curves in Fig.~\ref{fig:fpr_curves} for the reversed case, the DRW realization case, and the DRW realization case with a positive-only search range for comparison. The reversing procedure generates a monotonic FPR curve (red line in Fig.~\ref{fig:fpr_curves}) decreasing to small values for $\rm BF_{lag} > 10$. In contrast, the two DRW-based FPR curves (black lines in Fig.~\ref{fig:fpr_curves}) fluctuate near FPR$\sim0.6$ before plunging to low values for $\rm BF_{lag} \gtrsim 10$. The cumulative $\rm BF_{lag}$ distributions in the DRW case are under-sampled at high $\rm BF_{lag}$ due to the small number of reliably recovered lags, especially the spurious ones. This under-sampling causes jumps and artificial features at high $\rm BF_{lag}$. Using multiple DRW realizations per lag would mitigate bias in the distribution from coincidence and greatly increases the number of accepted spurious lags, especially at high $\rm BF_{lag}$, providing a more robust FPR estimate. For our simple comparison, we note that generally the reverse-case FPR curve resembles a smoothed version of the DRW-based FPR curves, and therefore adopt the reverse-case FPR curve for all lags. The FPR curve is more sensitive to the data quality, temporal baselines, and data overlap between the two light curves and less sensitive to the search range used in finding the lags.
% \blue{($\approx40$, with two having $\rm BF_{lag} > 1$)}

Since the number of lags measured is low ($\sim 20$), we cannot split this FPR estimation procedure by emission line, as in \citetalias{MacDougallEtAl2026}. This may bias the FPR estimates if the $\rm BF_{lag}$ distribution is dominated by one emission line over the others. Indeed, there are more potential lags for \MgII\ than for the other lines. In this way, our \hbeta\, lag FPRs assume a mostly \MgII-based $\rm BF_{lag}$ distribution. Future BHM-RM lag analysis will be able to address such issues. 

\subsection{Method Comparison}\label{subsec:litmus_comp}

Here we briefly compare results between PyROA and LITMUS. For a full comparison between all methods, see Appendix~\ref{app:lag_comp}.

In Table~\ref{tab:sample}, of the 20 lags recovered with PyROA, only four are recovered by LITMUS with an FPR $< 33\%$ (using a symmetric lag search). A detailed comparison (Appendix~\ref{app:lag_comp}) shows that with the current BHM data and a symmetric lag range, LITMUS is more affected by aliasing than PyROA, producing a poorer lag recovery rate. Nonetheless, the PyROA and LITMUS lags for these four are consistent with each other within $2\sigma$, although a larger sample is needed for a proper comparison. Utilizing the positive-only search range results adds one more lag (COS-479, \MgII) with a bronze ranking. While the relative length of the continuum and response light curves on the recovery of lags needs to be studied in further detail, the $2\sigma$ consistency between both methods in four lags suggests our quality cuts may work for a wide variety of data.
% \red{So LITMUS misses most of the PyROA lags when the search range is symmetric, which means it is somehow more prone to negative lag aliases? }

For completeness, Table~\ref{tab:sample} includes the eight additional lags recovered from the LITMUS symmetric lag search (with \citetalias{MacDougallEtAl2026} FPR $<33\%$) not found in the PyROA search. We perform the same \citetalias{ShenEtAl2024}-based FPR analysis and ranking using the LITMUS lag. None of these lags are ranked gold using the \citetalias{ShenEtAl2024} method, and only one is ranked gold from the \citetalias{MacDougallEtAl2026} method. The rest of these lags are mostly bronze-ranked and use \CIV, less reliable than those in the original 20-lag sample. From visual inspection, the lag posteriors are generally of worse quality than PyROA's, often falling within seasonal gaps or near the lag search boundary, or lacking a prominent peak.

Assuming a positive-only lag search with LITMUS adds an additional 31 lags to our original PyROA-selected sample, including the aforementioned eight. These lags are of similar quality or worse than the eight lags recovered with a symmetric lag search. Only a few of these lags are rated gold or silver with either ranking method. With a positive-only search range, lag methods are more likely to recover ``reliable" lags that are positive at the $2\sigma$ level coinciding with aliased peaks. Therefore, we opt to exclude these in our final results as future work is required to test bias from lag range assumptions using these four lag estimation techniques. Working with \MgII, \citet{YuEtAl2023} suggest using a combination of symmetric and asymmetric lag search results. In our case, the symmetric results present a more stringent cut on the recovered LITMUS lags than the asymmetric ones, restricting our sample to the current 28 even if we include the asymmetric results. 
% An in-depth comparison between LITMUS and other lag methods is given in Appendix~\ref{app:lag_comp}.

As the FPRs for individual objects are tied to the ranking they receive from the \citetalias{MacDougallEtAl2026} method, the FPRs within a given rank are mostly consistent. The \citetalias{ShenEtAl2024} FPRs can vary from 0--30\% within a given rank. Considering the eight lags recovered only by LITMUS, the \citetalias{ShenEtAl2024} method generally underestimates the FPR relative to the \citetalias{MacDougallEtAl2026} method. In this case, lag recovery is poor in general as most lags received a bronze rank. A true comparison necessitates using the PyROA model instead of PyCCF for \citetalias{ShenEtAl2024}-FPR estimation.

From our analysis, LITMUS and PyROA are able to recover lags from different sub-populations of the full sample. Either 1) the statistical implementation of PyROA and LITMUS lend themselves better to different types/qualities of data or 2) the objects found by one and not the other have high intrinsic FPRs, underestimated by current methods. Judging by the \citetalias{ShenEtAl2024} criteria applied to both samples, PyROA seems to recover higher-ranked lags on the whole. However, because the overlap between these two populations is small, a larger sample with a higher-quality dataset is required for a more thorough investigation. It may be that the procedure from \citetalias{ShenEtAl2024} is most restrictive, and will recover only the most secure lags. We note that this dataset with only six months of spectroscopy is of relatively poor length for detecting lags. Longer datasets are needed to verify the comparison between these methods.
% , detrimental for most sample analysis from the introduction of bias

\subsection{Comparison to SDSS-RM}
Overall, our procedure generates a lag recovery rate of 28/184 ($\approx$15\%) for lags with $\tau_{\rm exp}$ less than the spectroscopic baseline. The lag recovery rate is (17, 24, 8, 18)\% for (\hbeta, \halpha, \MgII, \CIV). Six-month spectroscopy from SDSS-RM in 2014 with a similar spectroscopic and photometric cadence produced a recovery rate of $\sim9\%$ for \hbeta\ and $\sim6\%$ for \MgII\ \citep{ShenEtAl2016}. Their recovery involved only PyCCF, with quality cuts of light curve SNR $>10$ and lag positivity at the $1\sigma$ level. However, we note that they arbitrarily choose only the 15/100 best lags, biasing their recovery rate. They also use the spectroscopic continuum as a reference light curve instead of independent photometry.

\citet{GrierEtAl2017} report \halpha\ and \hbeta\ lags with six-month SDSS-RM spectroscopy and external photometry, again with a similar cadence to ours. Their results, using JAVELIN and CREAM \citep{cream}, provide a recovery rate of 19\% for \hbeta\ and 33\% for \halpha. These results use the same weighting scheme as used in PyPETaL and similar ranking criteria as in Section~\ref{subsec:lag_reliability}. Our recovery rates are similar, slightly worse for \halpha\ lags. Considering our data can only provide a maximum of 25 \halpha\ lags, increasing our recovery rate to 33\% would require only two additional recovered lags. Such a difference is well within what is expected from different datasets.

If we utilize only JAVELIN instead of PyROA for a proper comparison, we find recovery rates of (14, 16, 16, 28)\% for (\hbeta, \halpha, \MgII, \CIV). Thus, JAVELIN produces similar recovery rates to PyROA for the Balmer lines, but offers modest improvements for \MgII\ and \CIV. Inspection of our JAVELIN \MgII+\CIV\ posteriors shows obvious signs of aliasing and clustering near the edge of the lag search range, moreso than PyROA, which inflates the lag recovery rates.

\subsection{$M_{\rm BH}$ in COSMOS}

\input{Table1.txt}

We compare our 28 obtained lags to $R-L$ relations from the literature in Fig.~\ref{fig:rl_relation}. As expected from our choice of search range, our lags lie close to the given relations, especially \halpha\, and \hbeta. This is clearly seen in \halpha, where estimates more closely follow the \citet{BentzEtAl2013} relation, systematically lower than the \citetalias{ShenEtAl2024} relation. We expect these Balmer lags to be the most secure, as the strong narrow \OIII\ emission lines near \hbeta\ are available to serve as an internal calibrator for the spectro-photometry. Furthermore, the Balmer lines have shown to be the most responsive and to exhibit shorter lags with minimal scatter \citep{MacDougallEtAl2025, ShenEtAl2024}. 

Our Balmer-line lags exhibit low scatter with respect to the $R-L$ relation ($\sim$0.1~dex), while the scatter for \MgII\ is much larger ($\sim$0.3~dex). While the \MgII\ lags still lie near the $R$-$L$ relation, their larger scatter likely indicates poorer lag quality. We measure only two secure \CIV\, lags with PyROA, both of which align with the \citet{MacDougallEtAl2025} relation. The six LITMUS \CIV\, lags lie offset from the \citet{MacDougallEtAl2025} estimate, closer to that measured in \citetalias{ShenEtAl2024}. More work must be done to determine the cause of the difference in these two relations, which may be related to survey depth, data quality, and seasonal gap aliasing. Indeed, most of these higher-than-expected \CIV\, lags are of lower quality. 

\begin{figure}
    \centering
    \includegraphics[width=0.5\textwidth]{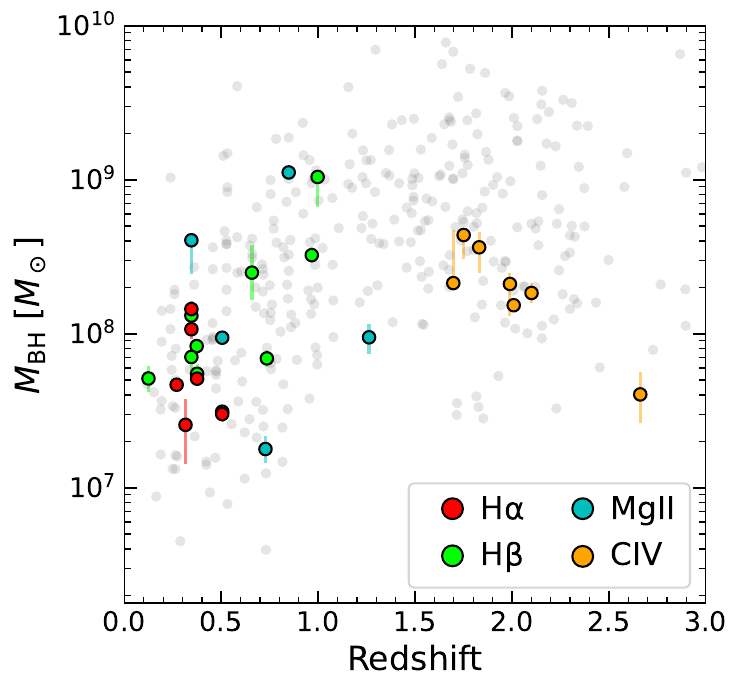}
    \caption{Our \mbh\ estimates as a function of redshift, color-coded by emission line. Displayed uncertainties do not include the $\sim0.3$~dex scatter in the virial factor. SDSS-RM results \citepalias{ShenEtAl2024} are shown as gray dots.}
    \label{fig:mbh}
\end{figure}

As in \citetalias{ShenEtAl2024}, we estimate the line dispersion $\sigma_{\rm line, rms}$ from the RMS spectrum of the fitted PrepSpec model. PrepSpec innately subtracts the continuum from the line beforehand, removing any potential contamination in the RMS spectrum which will ultimately decrease $\sigma_{\rm line, rms}$. We assume the average virial factor from \citetalias{ShenEtAl2024} of $\log_{10}(f) = 0.62$, obtained utilizing dynamical modeling estimates for a few dozen local sources. We note that this virial factor is an ensemble value derived from a local sample. In reality, $f$ differs between objects and depends on the geometry and kinematics of each AGN's BLR for each line. For example, for a BLR with Keplerian disk kinematics, the virial factor depends on inclination as $\propto 1/\sin(i)$ \citep{BentzEtAl2022}. Future work is needed to determine the effect of using this local, ensemble virial factor on diverse populations. Nonetheless, adopting the above average prescription along with our rest-frame time lags yields 28 \mbh\, estimates spanning $10^{7-9} M_{\odot}$ at redshifts $z \lesssim 2.5$ (Table~\ref{tab:sample} and Fig.~\ref{fig:mbh}). 

We compile all relevant variability and lag information for the full 184 lag sample, across methods, in a publicly available catalog \citep{zenodo_catalog}, described in Table~\ref{tab:catalog}. It should be noted that the reported \mbh\ estimates do not include the virial factor uncertainty of 0.32~dex. This uncertainty dominates the lag- and velocity-based uncertainties for most estimates, with the entire sample having an ``intrinsic" \mbh\ uncertainty of $\lesssim0.1$~dex on average.

% \red{Might be good to show a Mbh versus redshift plot, color-coded by line used for lag and mass. }

% \begin{itemize}
    % \item Compare SNR2 and the Bayes factors from \citet{MacDougallEtAl2026}
    % \item Define a gold, silver, and bronze sample in a similar way to \citet{ShenEtAl2024} and \citet{MacDougallEtAl2026}
    % \item All lags where $\tau_{\rm exp} > $ spectroscopic baseline are ranked bronze
    % \item Compare our sample in $R-L$ plot to prior results
    % \item Compare LITMUS to PyROA
    % \item Compare using symmetric to asymmetric prior for LITMUS
    % \item Compare LITMUS scoring to our scoring
% \end{itemize}

\section{Conclusion}\label{sec:conclusion}

We present the first reverberation mapping measurements from the SDSS-V BHM-RM project in the COSMOS field, using six months of reliable plate-based spectroscopy (late 2020 to mid-2021) and ZTF+DECam photometry. Utilizing a variety of lag-estimation methods and procedures to define lag reliability, we obtain 28 \mbh\ estimates for 22 unique broad-line AGNs ($10^{7-9} M_{\odot}$) at $z \lesssim 2.5$ using \halpha, \hbeta, \MgII, or \CIV. Combining procedures from prior SDSS-RM and OzDES analyses grants a larger lag success rate, as opposed to using either individually. For our dataset in particular, the SDSS-RM procedure is the most restrictive, identifying only the highest-quality time lags, while the OzDES procedure recovers more time lags in total, including those of lower quality. Both methods produce BLR size estimates consistent with each other and literature $R-L$ relations.

While reliable SDSS-V BHM data using plates to position the optical fibers last until mid-2021, improvements in calibration of spectra taken later with fiber positioning robots will enable analysis up to late 2026. Such data will improve \mbh\ constraints and increase the size and diversity of our sample, especially for \MgII\ and \CIV\ measurements. SDSS-V BHM \mbh\ estimates enable cross-facility and multi-wavelength analysis, especially in COSMOS, XMM-LSS, and CDF-S. The LSST Deep-Drilling Fields will observe optical photometric variability over a decade with a $\sim$daily cadence, and Roman will measure $\sim$10d-cadence near-infrared spectroscopic and photometric variability over its five-year survey. COSMOS is also a Euclid Auxiliary Field, visited $\sim$4 times by the end of Euclid's five-year survey \citep{Euclid_EAF}, and the focus of a number of JWST programs \citep[e.g.,][]{cosmos3d}. Complementary analysis across all surveys in the coming years will provide tight constraints on AGN variability, BLR structure, and SMBH properties across redshift and accretion state.

\begin{acknowledgments}
This work is partially supported by NSF grant AST-2509424 (ZS and YS). FEB acknowledges support from ANID-Chile BASAL CATA FB210003 and FONDECYT Regular 1241005. LHG acknowledges financial support from ANID program FONDECYT Iniciaci\'on 11241477.

Funding for the Sloan Digital Sky Survey V has been provided by the Alfred P. Sloan Foundation, the Heising-Simons Foundation, the National Science Foundation, and the Participating Institutions. SDSS acknowledges support and resources from the Center for High-Performance Computing at the University of Utah. SDSS telescopes are located at Apache Point Observatory, funded by the Astrophysical Research Consortium and operated by New Mexico State University, and at Las Campanas Observatory, operated by the Carnegie Institution for Science. The SDSS web site is \url{www.sdss.org}.

SDSS is managed by the Astrophysical Research Consortium for the Participating Institutions of the SDSS Collaboration, including the Carnegie Institution for Science, Chilean National Time Allocation Committee (CNTAC) ratified researchers, Caltech, the Gotham Participation Group, Harvard University, Heidelberg University, The Flatiron Institute, The Johns Hopkins University, L'Ecole polytechnique f\'{e}d\'{e}rale de Lausanne (EPFL), Leibniz-Institut f\"{u}r Astrophysik Potsdam (AIP), Max-Planck-Institut f\"{u}r Astronomie (MPIA Heidelberg), Max-Planck-Institut f\"{u}r Extraterrestrische Physik (MPE), Nanjing University, National Astronomical Observatories of China (NAOC), New Mexico State University, The Ohio State University, Pennsylvania State University, Smithsonian Astrophysical Observatory, Space Telescope Science Institute (STScI), the Stellar Astrophysics Participation Group, Universidad Nacional Aut\'{o}noma de M\'{e}xico, University of Arizona, University of Colorado Boulder, University of Illinois at Urbana-Champaign, University of Toronto, University of Utah, University of Virginia, Yale University, and Yunnan University.

ZTF is supported by the National Science Foundation and a collaboration including current partners Caltech, IPAC, the Oskar Klein Center at Stockholm University, the University of Maryland, the University of California at Berkeley, the University of Wisconsin at Milwaukee, University of Warwick, Ruhr University, Cornell University, Northwestern University and Drexel University. Operations are conducted by COO, IPAC, and UW. The ztfquery code was funded by the European Research Council (ERC) under the European Union's Horizon 2020 research and innovation programme (grant agreement n°759194 - USNAC, PI: Rigault).

This project used data obtained with the Dark Energy Camera (DECam), which was constructed by the Dark Energy Survey (DES) collaboration. Funding for the DES Projects has been provided by the US Department of Energy, the U.S. National Science Foundation, the Ministry of Science and Education of Spain, the Science and Technology Facilities Council of the United Kingdom, the Higher Education Funding Council for England, the National Center for Supercomputing Applications at the University of Illinois at Urbana-Champaign, the Kavli Institute for Cosmological Physics at the University of Chicago, Center for Cosmology and Astro-Particle Physics at the Ohio State University, the Mitchell Institute for Fundamental Physics and Astronomy at Texas A\&M University, Financiadora de Estudos e Projetos, Funda\c{c}\~a Carlos Chagas Filho de Amparo \`a Pesquisa do Estado do Rio de Janeiro, Conselho Nacional de Desenvolvimento Cient\'ifico e Tecnol\'ogico and the Minist\'erio da Ci\^{e}ncia, Tecnologia e Inova\c{c}\~ao, the Deutsche Forschungsgemeinschaft and the Collaborating Institutions in the Dark Energy Survey.

The Collaborating Institutions are Argonne National Laboratory, the University of California at Santa Cruz, the University of Cambridge, Centro de Investigaciones En\'ergeticas, Medioambientales y Tecnol\'ogicas–Madrid, the University of Chicago, University College London, the DES-Brazil Consortium, the University of Edinburgh, the Eidgen\"{o}ssische Technische Hochschule (ETH) Z\"{u}rich, Fermi National Accelerator Laboratory, the University of Illinois at Urbana-Champaign, the Institut de Ci\`encies de l’Espai (IEEC/CSIC), the Institut de F\'isica d’Altes Energies, Lawrence Berkeley National Laboratory, the Ludwig-Maximilians Universit\"{a}t M\"{u}nchen and the associated Excellence Cluster Universe, the University of Michigan, NSF NOIRLab, the University of Nottingham, the Ohio State University, the OzDES Membership Consortium, the University of Pennsylvania, the University of Portsmouth, SLAC National Accelerator Laboratory, Stanford University, the University of Sussex, and Texas A\&M University.

\end{acknowledgments}

\facilities{APO}

\software{
Astropy \citep{astropy1, astropy2, astropy3},
celerite \citep{celerite},
JAVELIN \citep{javelin1, javelin2},
LITMUS \citep{litmus},
Matplotlib \citep{matplotlib},
NumPy \citep{numpy},
PrepSpec \citep{ShenEtAl2016},
PyCALI \citep{pycali},
PyCCF \citep{pyccf},
PyPETaL \citep{pypetal},
PyROA \citep{pyroa},
SciPy \citep{scipy},
ztfquery \citep{ztfquery}
}

\bibliography{Refs}{}
\bibliographystyle{aasjournal}

\appendix

\begin{table*}
\centering
\caption{Sample Catalog Description\label{tab:fields}}
\begin{tabular}{|l|l|c|l|}
    \hline
    \textbf{Column Name} & \textbf{Data Type} & \textbf{Unit} & \textbf{Description} \\
    \hline
    NAME & STR & & BHM-RM source name \\
    RA & DOUBLE & deg & Right Ascension (J2000) \\
    DEC & DOUBLE & deg & Declination (J2000) \\
    Z & DOUBLE & & Spectroscopic redshift \\
    LINE & STR & & Emission line (\halpha=ha, \hbeta=hb, \MgII=mg2, \CIV=c4) \\
    \hline
    \tablenotemark{$\bigstar$}MBH & DOUBLE(3) & $M_{\odot}$ & RM-based $M_{\rm BH}$, excluding scatter in the virial factor \citepalias[0.32~dex;][]{ShenEtAl2024} \\
    LOGLUM & DOUBLE & erg s$^{-1}$ & $\log_{10}(\lambda L_{\lambda})$ at $\rm 1350\AA$ (\CIV), $\rm 3000\AA$ (\MgII),  $\rm 5100\AA$ (\halpha+\hbeta) \\
    \tablenotemark{$\bigstar$}SIGMA0\_CONT & DOUBLE(3) & & Fractional intrinsic rms variability \citep[$\sigma_0$;][]{ShenEtAl2019} for the continuum \\
    SNR2\_CONT & DOUBLE & & SNR2 \citep{ShenEtAl2019} for the continuum \\
    \tablenotemark{$\bigstar$}SIGMA0\_LINE & DOUBLE(3) & & Fractional $\sigma_0$ for the line \\
    SNR2\_LINE & DOUBLE & & SNR2 for the line \\
    DISP & DOUBLE & km s$^{-1}$ & Line dispersion $\sigma_{\rm line, rms}$ measured from the RMS spectrum \\
    DISP\_ERR & DOUBLE & km s$^{-1}$ & Uncertainty in DISP \\
    \hline
    \tablenotemark{$\bigstar$}\tablenotemark{$*$}LAG\_xxx & DOUBLE(3) & day & Measured observed lag \\
    \tablenotemark{$*$}RMAX\_xxx & DOUBLE & & Maximum of the ICCF within $\pm1\sigma$ of the observed lag ($r_{\rm max}$) \\
    \tablenotemark{$*$}FRAC\_REJECTED\_xxx & DOUBLE & & Fraction of the posterior rejected from the weighting procedure \\
    \tablenotemark{$*$}RELIABLE\_xxx & BOOL & & Whether the lag was reliable: passing the three \citetalias{ShenEtAl2024} criteria \\ 
    &&& and having $\tau_{\rm exp}$ less than the spectroscopic baseline \\
    \tablenotemark{$\bullet$}FRAC\_EVIDENCE\_xxx & DOUBLE & & Fraction of the LITMUS posterior evidence $Z$ in the primary peak \\
    \tablenotemark{$\bullet$}BF\_LAG\_xxx & DOUBLE & & $\rm \log_{10}(BF_{lag})$ in the primary peak \\
    \tablenotemark{$\bullet$}BF\_STRUC\_RESP\_xxx & DOUBLE & & $\rm \log_{10}(BF_{struc,resp})$ in the primary peak \\
    \tablenotemark{$\bullet$}BF\_STRUC\_CONT\_xxx & DOUBLE & & $\rm \log_{10}(BF_{struc,cont})$ in the primary peak \\
    FPR\_S24 & DOUBLE & \% & False-positive rate from the \citetalias{ShenEtAl2024} method \\
    \tablenotemark{$\bullet$}FPR\_M26\_xxx & DOUBLE & \% & False-positive rate from the \citetalias{MacDougallEtAl2026} method \\
    RANK\_S24 & STR & & Rank given from the \citetalias{ShenEtAl2024} method (\bronze, \silver, or \gold) \\
    \tablenotemark{$\bullet$}RANK\_M26\_xxx & STR & & Rank given from the \citetalias{MacDougallEtAl2026} method (\bronze, \silver, or \gold) \\
    \hline
\end{tabular}\label{tab:catalog}
\vspace{1pt}
\begin{adjustwidth}{0cm}{}
    $^\bigstar$These columns have the value and uncertainties in the following order: [lower error, value, upper error]\\[2pt]
    $^*$These are four columns, for each lag estimated, where xxx is: PYCCF, JAVELIN, PYROA, LITMUS\_SYM, or LITMUS\_ASYM.\\[2pt]
    $^\bullet$For LITMUS results, these are two columns for each lag search range (symmetric/asymmetric), where xxx is: SYM or ASYM.\\[2pt]
    Our final estimates use RELIABLE\_PYROA=TRUE for PyROA lags, and (RELIABLE\_LITMUS\_SYM=TRUE) \& (FPR\_M26\_SYM$<$=0.33) for LITMUS lags. If a given source/line has both lags, we utilize the PyROA lag.
\end{adjustwidth}
\end{table*}

\section{Lag Quality}\label{app:reliability}

\begin{figure}
    \centering
    \includegraphics[width=\linewidth]{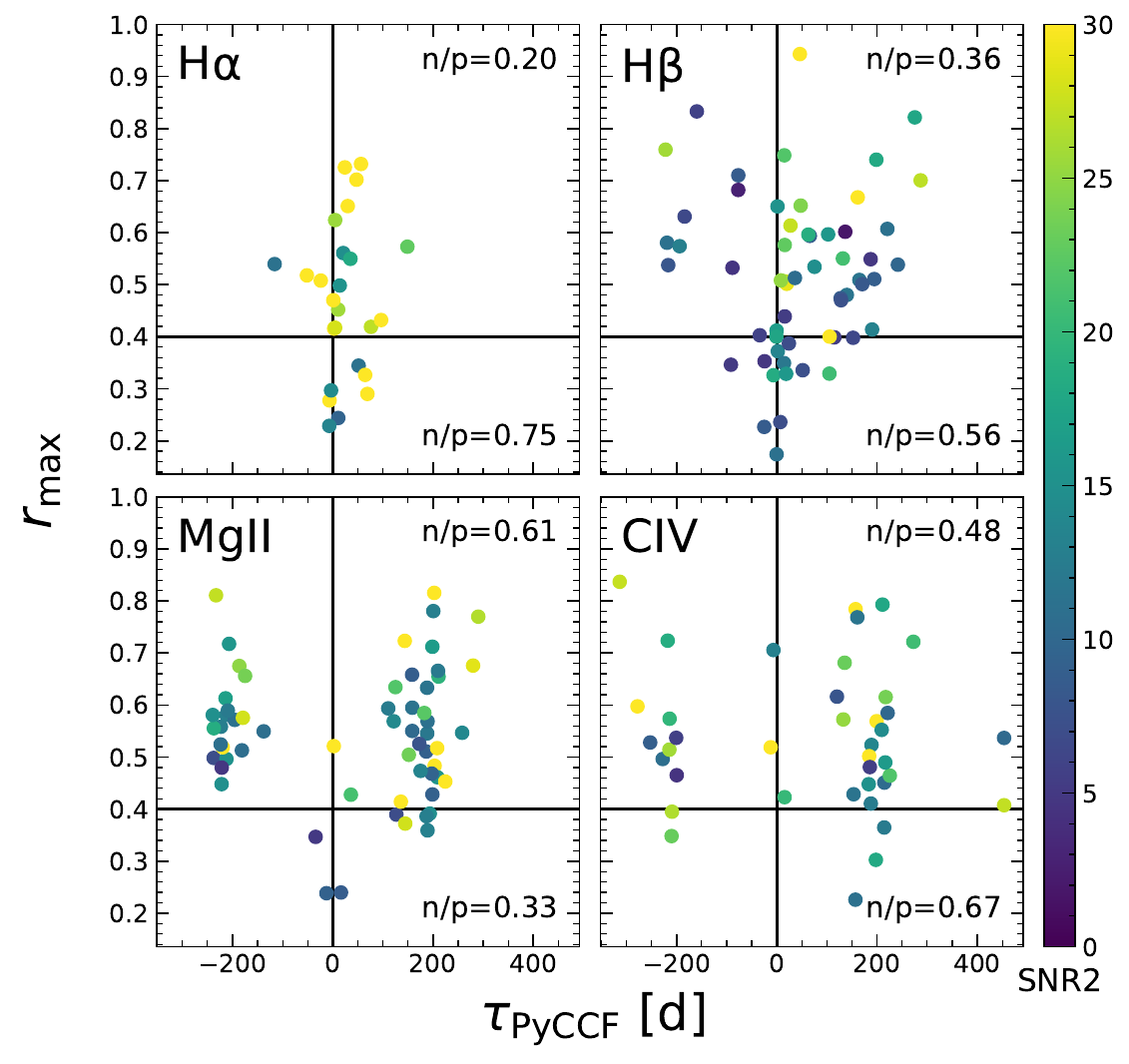}
    \caption{Lag detection reliability for the full sample using PyCCF, for each line. Lags are colored by their line light curve SNR2. For each line, the ratio of negative to positive lags is shown above and below the $r_{\rm max}=0.4$ threshold.}
    \label{fig:np_ratio}
\end{figure}

We investigate the overall quality of our lags in Fig.~\ref{fig:np_ratio}. This shows the PyCCF lags as a function of $r_{\rm max}$ and SNR2, an adjusted $\chi^2$ statistic for intrinsic variability \citep{ShenEtAl2019}. For reference, the sample from \citetalias{ShenEtAl2024} shows a low negative-to-positive lag ratio at $r_{\rm max} > 0.4$, and a higher one at $r_{\rm max} \leq 0.4$. This indicates that lags passing the $r_{\rm max}$ threshold have a lower false-positive rate. 

Because of the length of our data, the negative-to-positive lag ratio is still fairly high above the $r_{\rm max}$ threshold. These ratios are moderate for \halpha\ and \hbeta, but substantial ($\gtrsim 0.5$) for \MgII\ and \CIV. In fact, the \MgII\ ratio is higher for lags below the threshold than above the threshold. The \MgII\ and \CIV\ distributions are bimodal, with lags clustering near the spectroscopic baseline. This is also somewhat apparent to a much lesser extent for \hbeta. 

This indicates that the \halpha\ and \hbeta\ lags are decently reliable, while the \MgII\ and \CIV\ lags have high FPRs, as seen from the low number of \MgII\ and \CIV\ results in the final lag sample. Indeed, it is difficult to measure longer lags at $z > 1$ with only six months of spectroscopy. SDSS-RM simulations produce a \hbeta\ lag false-positive rate $\lesssim20\%$ with six months of spectroscopy \citep{ShenEtAl2015}, also seen in their first-year \halpha\ and \hbeta\ results \citep{GrierEtAl2017}. If this holds for our entire sample, only $\sim$5 lags would be false-positives, likely the bronze-ranked estimates. The higher negative-to-positive ratio in our sample indicates a \hbeta\ FPR $\sim$35\%, bringing the number of artificial \hbeta+\halpha\ lags to $\sim$5/16. Indeed, many of the lags have individual FPRs $\sim$30\%. We note our stringent quality cuts mitigate such cases. This inflated sample-wide FPR may be due to the presence of seasonal gaps in the photometry, or depth of the included ZTF data. Increased photometric scatter from ZTF in fainter sources may create more difficulty in measuring reliable lags.

\citetalias{ShenEtAl2024} find that the upper right quadrant of each panel are populated with light curves displaying large SNR2. This suggests that sources with high-quality light curves that reveal intrinsic broad-line variability will produce reliable lags. While the number of \halpha\ lags is small, most lags with $r_{\rm max} > 0.4$ display a high SNR2 $> 30$. Furthermore, these high-SNR2 lags are largely positive. These trends indicate that the \halpha\ data is able to capture intrinsic variability and produce reliable lags. The \hbeta\ lags also display a trend of increased SNR2 for positive lags with $r_{\rm max} > 0.4$, albeit with lower SNR2 on average and more low-SNR2 contaminants. This is likely due to the difficulty in measuring broad \hbeta\ relative to \halpha, and also represented in the higher negative-to-positive ratio. Due to the significant impact of aliasing in \MgII\ and \CIV, most lags at $r_{\rm max} > 0.4$ cluster near the spectroscopic baseline with varying SNR2.

\section{Detailed Method Comparison}\label{app:lag_comp}

% \begin{figure*}
%     \centering
%     \includegraphics[width=\linewidth]{figures/lag_comp_all.pdf}
%     \caption{A comparison of lags measured with LITMUS to both PyROA (\emph{top}) and PyCCF (\emph{bottom}). The left column shows LITMUS lags with a symmetric lag search range, while the right column uses a positive only search range. Colors correspond to whether the PyROA/PyCCF lags are reliable, and shapes correspond to whether the LITMUS lag is reliable. A dashed line shows where the two lags are equal.}
%     \label{fig:lag_comp}
% \end{figure*}

\begin{figure*}
    \centering
    \includegraphics[width=0.8\linewidth]{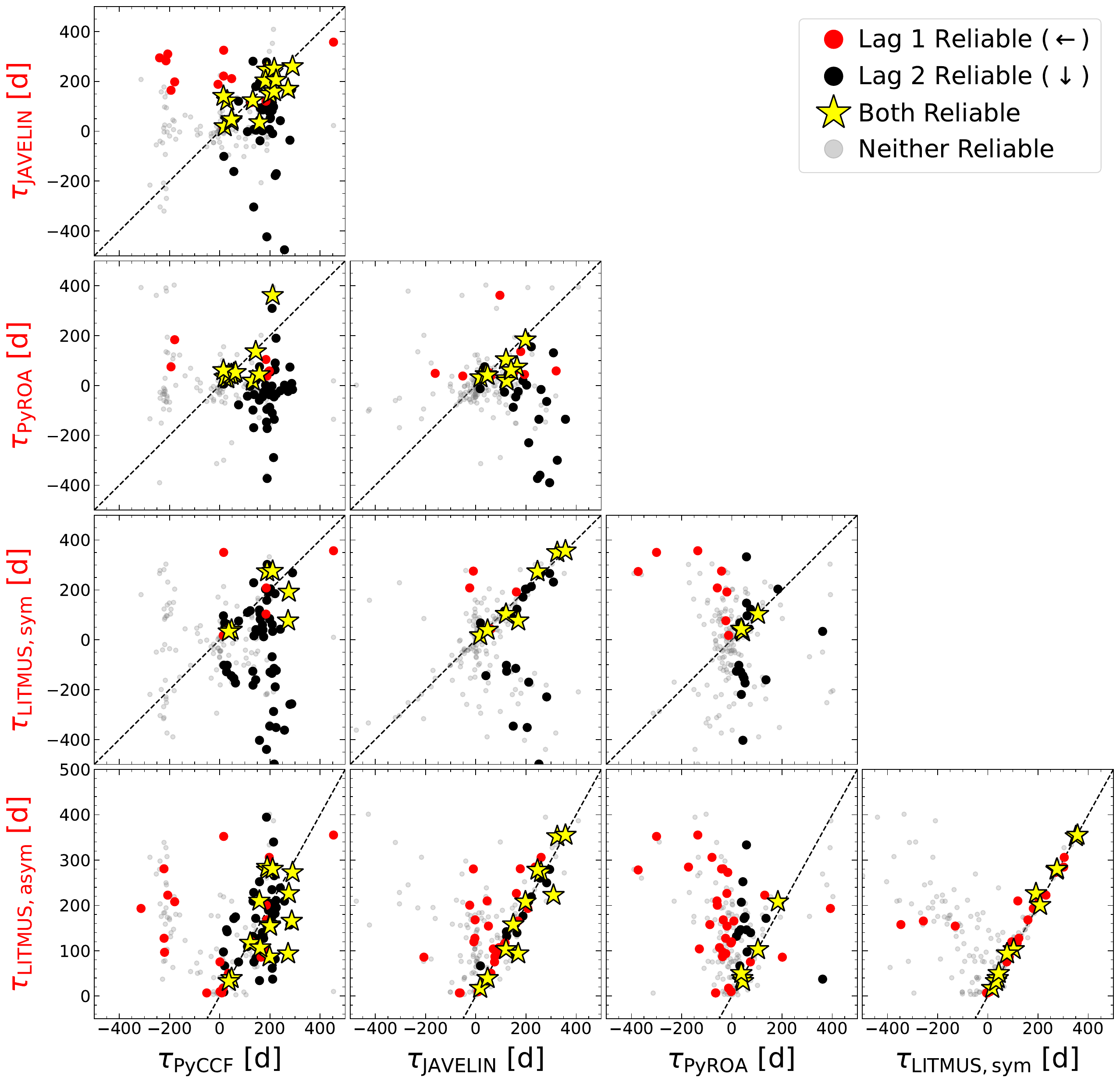}
    \caption{A comparison of lags from all considered methods, including two different LITMUS estimates from assuming different lag search ranges (symmetric or asymmetric). Red circles show lag estimates that are reliable (i.e., passing the three criteria from Section~3.1) for the first method (y-axis), and black circles show reliable lags from the second method (x-axis). Yellow stars represent estimates considered reliable from both methods, while gray circles show estimates that were not considered reliable from either method. A dashed line shows where the two lags are equal.}
    \label{fig:lag_comp}
\end{figure*}

Fig.~\ref{fig:lag_comp} compares lag estimates derived from all methods to each other, with LITMUS estimates assuming either a symmetric or asymmetric (i.e., positive-only) lag range. For this comparison, we consider lags reliable if they pass the three criteria adopted from \citetalias{ShenEtAl2024} (i.e., positivity by $2\sigma$, $r_{\rm max} > 0.4$, $<50\%$ of the posterior discarded), prior to visual inspection. LITMUS lags assume an additional constraint of having a FPR $< 33\%$.

As found in Appendix~\ref{app:reliability}, PyCCF identifies a number of aliased lags longer than the spectroscopic baseline, clustering at $\sim200$d timescales. PyCCF recovered a larger number of lags within this aliased region than the other methods, although JAVELIN also recovers many. This echoes the SDSS-RM finding that JAVELIN and PyCCF are susceptible to aliasing from seasonal gaps. Notably, PyROA discards most of these lags, with only a few being declared reliable in the aliasing region $\gtrsim$200d or the negated aliasing region near $-$200d. The lags declared reliable by both PyCCF and PyROA, as well as by both JAVELIN and PyROA, all fall near the one-to-one relation, with a scatter of $\sim$tens of days. These lags also largely lie lower than the spectroscopic baseline. Thus, as in \citetalias{ShenEtAl2024}, we find that PyROA is the most restrictive and produces the most reliable, least aliased lags.

We first focus on the LITMUS lags recovered with a symmetric lag range (row three of Fig.~\ref{fig:lag_comp}). LITMUS performs better than PyCCF, discarding most of the PyCCF-recovered lags in the aliasing region. The reliable lags it shares with PyCCF lie mostly larger than the spectroscopic baseline, with only a few short lags identified. Seeing as the LITMUS model we adopt is fairly similar to the JAVELIN model, it is unsurprising that LITMUS performs similarly to JAVELIN. Interestingly, LITMUS rejects a number of spurious aliased lags found by JAVELIN, picking up a few that JAVELIN declared as having a lag near zero. Nonetheless, the lags recovered by both JAVELIN and LITMUS lie very closely to the one-to-one relation. In this way, LITMUS may serve as a more robust alternative to JAVELIN, and PyCCF, removing a modest number of aliased lags.

As stated in Section~3.3, only four of the lags recovered by LITMUS are also recovered by PyROA. PyROA rejects all of the aliased LITMUS lags larger than the spectroscopic baseline, and a few shorter ones. All of these additional LITMUS lags, not recovered by PyROA, are included in our final results, but we note these two shorter lags (COS-762 \& COS-837) are the most reliable. While LITMUS seems to be more restrictive than PyCCF or JAVELIN, it still recovers a significant number of aliased lags at large timescales. Thus, with the quality and length of our dataset, LITMUS is more affected by aliasing than PyROA. Additional restrictions on the quality of lags may be able to bring performance closer to PyROA's, such as including a limit on the Bayesian metrics $\rm BF_{struc, cont}$ or $\rm BF_{struc, resp}$.

We now focus on LITMUS lags using an asymmetric lag range (row four of Fig.~\ref{fig:lag_comp}), as used in \citetalias{MacDougallEtAl2026}. The asymmetric lag range is less restrictive in terms of lag recovery, producing 36 total as opposed to a symmetric lag range producing 12. A positive only lag search enables more estimates to pass one of the three criteria -- positivity by $2\sigma$. LITMUS is as susceptible to aliasing either with an asymmetric lag range or a symmetric one. Nearly 33\% (13/36) of reliable lags lie above 200d with an asymmetric search range, while 40\% (5/12) lie above 200d with a symmetric search range. Most of these lags coincide with PyCCF lags lying within the aliasing region ($\sim\pm200$d). As expected, assuming an asymmetric lag range enables recovery of more lags, compared to the symmetric case. We note that a significant fraction of LITMUS lags are consistent with each other regardless of the choice in lag search range.

The reliable asymmetric LITMUS lags are almost all consistent with the JAVELIN lags, moreso than when assuming a symmetric lag range. Additionally, more lags are declared reliable by both LITMUS and JAVELIN when assuming an asymmetric search range. Nearly 10 lags were recovered with the asymmetric LITMUS estimate, but measured to be near zero with JAVELIN. As many of these lags have poor posteriors after visual inspection, as stated in Section~3.3, they are likely aliased. Robust lag measurements require validation across methods, especially with PyROA. PyROA again only recovers four of the estimates found reliable by LITMUS, with many of the remaining LITMUS lags measured to be near zero with PyROA. Thus, compared to JAVELIN and PyROA, LITMUS with an asymmetric lag range is significantly affected by aliasing with our quality of data.

We emphasize that these ``reliable" lags have not been vetted through visual inspection, a key requirement of our lag reliability criteria. Most of the LITMUS and JAVELIN results passing the aforementioned three criteria have poor lag posteriors, with the peak being ill-formed (i.e., a multimodal distribution) or falling close to the edge of the lag search range. Lag recovery may also depend on the line used, which we do not consider here. \MgII\ lags in particular are difficult to measure due to its weak response \citep{GoadEtAl1993}. After careful consideration of all methods, PyROA is still the most resistant to aliasing, as in \citetalias{ShenEtAl2024}, and produces more reliable lag estimates regardless of the quality of the data. LITMUS (assuming a symmetric lag range) is mostly consistent with JAVELIN, but shows modest improvements in discarding aliased lags, albeit with poor quality posteriors.

If lags are recovered reliably from multiple methods, the estimates typically differ by a factor $\lesssim$1.5. Scatter between methods is smaller when comparing LITMUS and PyROA, and can reach $\sim$2.3 for some lags when comparing PyCCF/JAVELIN and PyROA. Given that the uncertainty in \mbh\ is dominated by the virial factor uncertainty $\approx$0.3~dex, and $R-L$ relations exhibit typical scatter $\sim$0.3~dex, the choice of method largely has little impact on subsequent analysis.

\end{document}

%% file: Table1.txt
\begin{deluxetable}{cllcccccccc}
\tablecaption{Sample Summary}\label{tab:sample}
\tablehead{
    \colhead{Name} & \colhead{R.A.} & \colhead{Decl.} & \colhead{$z$} & \colhead{Line} & \colhead{$\log_{10}(M_{\rm BH} / M_{\odot})$} & \colhead{$\rm \log_{10}( BF_{lag})$} & \multicolumn{2}{c}{FPR [\%]} & \multicolumn{2}{c}{Rank} \\
    & & & & & & & \colhead{S24} & \colhead{M26} & \colhead{S24} & \colhead{M26}
}
\startdata
\multicolumn{11}{c}{PyROA} \\
\hline
COS-911 & 149.872750 & 3.183770 & 0.12 & \hbeta & ${7.71}_{-0.09}^{+0.08}$ & \nodata & 17 & \nodata & \gold & \nodata \\[7pt]
COS-1330 & 150.346340 & 1.374526 & 0.32 & \halpha & ${7.41}_{-0.26}^{+0.17}$ & 0.51 & 16 & 18 & \silver & \bronze \\[7pt]
\multirow{ 2 }{*}{ COS-872 } & \multirow{ 2 }{*}{149.582810} & \multirow{ 2 }{*}{2.484319} & \multirow{ 2 }{*}{0.35} & \halpha & ${8.03}_{-0.06}^{+0.05}$ & \nodata & 24 & \nodata & \gold & \nodata \\
& & & & \hbeta & ${7.85}_{-0.10}^{+0.08}$ & \nodata & 14 & \nodata & \gold & \nodata \\[7pt]
\multirow{ 3 }{*}{ COS-552 } & \multirow{ 3 }{*}{149.761500} & \multirow{ 3 }{*}{2.318439} & \multirow{ 3 }{*}{0.35} & \halpha & ${8.16}_{-0.05}^{+0.05}$ & \nodata & 24 & \nodata & \gold & \nodata \\
& & & & \hbeta & ${8.12}_{-0.02}^{+0.02}$ & \nodata & 28 & \nodata & \gold & \nodata \\
& & & & \MgII & ${8.61}_{-0.22}^{+0.05}$ & \nodata & 32 & \nodata & \gold & \nodata \\[7pt]
COS-849 & 150.105200 & 1.981134 & 0.37 & \hbeta & ${7.92}_{-0.02}^{+0.02}$ & \nodata & 27 & \nodata & \gold & \nodata \\[7pt]
\multirow{ 2 }{*}{ COS-895 } & \multirow{ 2 }{*}{150.683140} & \multirow{ 2 }{*}{2.574634} & \multirow{ 2 }{*}{0.38} & \halpha & ${7.71}_{-0.04}^{+0.04}$ & 0.84 & 28 & 6 & \gold & \silver \\
& & & & \hbeta & ${7.74}_{-0.05}^{+0.06}$ & 1.24 & 27 & 0 & \silver & \gold \\[7pt]
\multirow{ 3 }{*}{ COS-479 } & \multirow{ 3 }{*}{149.598910} & \multirow{ 3 }{*}{1.209920} & \multirow{ 3 }{*}{0.50} & \halpha & ${7.48}_{-0.05}^{+0.04}$ & \nodata & 12 & \nodata & \gold & \nodata \\
& & & & \hbeta & ${7.49}_{-0.04}^{+0.04}$ & -0.06 & 15 & 46 & \gold & \nodata \\
& & & & \MgII & ${7.97}_{-0.03}^{+0.03}$ & -0.06 & 1 & 46 & \silver & \nodata \\[7pt]
COS-765 & 150.123660 & 2.358245 & 0.73 & \MgII & ${7.25}_{-0.09}^{+0.09}$ & \nodata & 8 & \nodata & \silver & \nodata \\[7pt]
COS-592 & 149.180700 & 2.604892 & 0.74 & \hbeta & ${7.84}_{-0.06}^{+0.05}$ & \nodata & 22 & \nodata & \gold & \nodata \\[7pt]
COS-505 & 150.558620 & 1.159679 & 0.85 & \MgII & ${9.05}_{-0.05}^{+0.04}$ & \nodata & 5 & \nodata & \silver & \nodata \\[7pt]
COS-842 & 150.462970 & 2.009011 & 0.97 & \hbeta & ${8.51}_{-0.05}^{+0.05}$ & \nodata & 10 & \nodata & \gold & \nodata \\[7pt]
COS-877 & 149.408520 & 2.325472 & 1.26 & \MgII & ${7.98}_{-0.11}^{+0.09}$ & \nodata & 9 & \nodata & \silver & \nodata \\[7pt]
COS-538 & 149.458240 & 1.565035 & 2.01 & \CIV & ${8.19}_{-0.03}^{+0.03}$ & 0.49 & 2 & 19 & \gold & \bronze \\[7pt]
COS-630 & 149.094260 & 2.783455 & 2.66 & \CIV & ${7.61}_{-0.19}^{+0.14}$ & \nodata & 23 & \nodata & \silver & \nodata \\[7pt]
\hline
\multicolumn{11}{c}{LITMUS} \\
\hline
COS-762 & 150.262560 & 2.670822 & 0.27 & \halpha & ${7.67}_{-0.05}^{+0.04}$ & -0.06 & 14 & 0 & \silver & \gold \\[7pt]
COS-823 & 150.633860 & 2.593712 & 0.66 & \hbeta & ${8.40}_{-0.17}^{+0.18}$ & -0.06 & 26 & 23 & \bronze & \bronze \\[7pt]
COS-659 & 150.814080 & 3.129070 & 1.00 & \hbeta & ${9.02}_{-0.20}^{+0.03}$ & 0.36 & 0 & 24 & \bronze & \bronze \\[7pt]
COS-837 & 149.367600 & 2.833984 & 1.70 & \CIV & ${8.33}_{-0.05}^{+0.34}$ & -0.29 & 24 & 30 & \silver & \bronze \\[7pt]
COS-770 & 149.533610 & 3.224064 & 1.75 & \CIV & ${8.64}_{-0.16}^{+0.04}$ & 0.49 & 5 & 32 & \silver & \bronze \\[7pt]
COS-843 & 150.158390 & 2.139555 & 1.83 & \CIV & ${8.56}_{-0.16}^{+0.10}$ & 0.10 & 0 & 31 & \bronze & \bronze \\[7pt]
COS-625 & 148.790600 & 2.496669 & 1.99 & \CIV & ${8.32}_{-0.21}^{+0.07}$ & -0.10 & 0 & 22 & \silver & \bronze \\[7pt]
COS-535 & 149.545330 & 1.668100 & 2.10 & \CIV & ${8.26}_{-0.07}^{+0.06}$ & 0.21 & 0 & 24 & \bronze & \bronze \\
\enddata
\end{deluxetable}